\documentclass[
  journal=pasa,
  manuscript=article-type,
  year=2020,
  volume=37,
]{cup-journal}
\usepackage{hyperref}
\hypersetup{
    colorlinks = true,
    allcolors  = {blue} 
}
\usepackage{amsmath}
\usepackage[nopatch]{microtype}
\usepackage{booktabs}
\usepackage{appendix}
\usepackage{amssymb}

\usepackage[T1]{fontenc}
\usepackage[normalem]{ulem}
\usepackage{xcolor}

\newcommand{\ms}{M$_{\odot}$}
\newcommand{\rs}{R$_{\odot}$}

\newcommand{\rev}[1]{\textbf{#1}} 
\renewcommand{\rev}[1]{#1} 

\title{Weakened Inspirals --- I: High Mass Ratio Common Envelope Interactions in RGB Stars}

\author{J. Nibbs}
\email[J. Nibbs]{jack.nibbs@hdr.mq.edu.au}
\affiliation{School of Mathematical \& Physical Sciences, Macquarie University, Macquarie Park, NSW 2109, Australia}
\alsoaffiliation{Astrophysics and Space Technologies Research Centre, Macquarie University, Macquarie Park, NSW 2109, Australia}

\author{Orsola De Marco}
\affiliation{School of Mathematical \& Physical Sciences, Macquarie University, Macquarie Park, NSW 2109, Australia}
\alsoaffiliation{Astrophysics and Space Technologies Research Centre, Macquarie University, Macquarie Park, NSW 2109, Australia}

\author{Lionel Siess}
\affiliation{Institut d’Astronomie et d’Astrophysique, Universit\'{e} Libre de Bruxelles (ULB, BLU), CP 226, B-1050 Brussels, Belgium}

\author{Ryosuke, Hirai}
\affiliation{RIKEN Cluster for Pioneering Research (CPR), RIKEN, Wako, Saitama 351-0198, Japan}
\alsoaffiliation{School of Physics and Astronomy, Monash University, VIC 3800, Australia}
\alsoaffiliation{OzGrav: The ARC Centre of Excellence for Gravitational Wave Discovery, Australia}

\author{Daniel J. Price}
\affiliation{School of Physics and Astronomy, Monash University, VIC 3800, Australia}
\alsoaffiliation{IPAG, Univ. Grenoble Alpes, CNRS, 38000 Grenoble, France}

\keywords{hydrodynamics - methods: numerical - stars: AGB and post-AGB - binaries: general} 

\begin{document}

\begin{abstract}
    Post-red and post-asymptotic giant stars in binary systems with main sequence companions, have periods in the range $\sim$50-2000~days, eccentricities as high as 0.6 and are surrounded by a circumbinary disc. Their orbital separations are small enough that the system must have gone through Roche lobe overflow when the primary was a full blown giant; Roche lobe overflow between a giant and a more compact companion tend to lead to a common envelope inspiral, leaving a circular orbit with periods much shorter than observed in these systems. In this first work in a series we explore to what extent a high mass ratio, $q \equiv M_2/M_1$, can lead to wider orbital separations, by carrying out a series of 3D, hydrodynamical CE binary interaction simulations with the smoothed particle hydrodynamics code \textsc{Phantom}. The giant is a 0.88~\ms, 90~\rs, red giant branch star and the companions have a range of masses such that $q =  0.68 - 1.5$. While larger $q$ values result in wider post-CE separations, the upper limit we predict is only 
    $\sim50$~\rs, smaller than the observed range. The pre-CE mass transfer phase is longer for  larger companion masses and around $q\gtrsim 1$ the nature of the CE inspiral changes significantly, showing greater stability,  as predicted by analytical theory. However, this phase is not converged with respect to simulation resolution and it is expected that a higher resolution would lead to even more stability and a longer pre-inspiral phase. Despite more material flowing through the $L_2$ and $L_3$ Lagrange points for higher $q$ values, with the potential for the formation of a circumbinary disc structure in this way, we conclude that, for our parameters, circumbinary discs are more likely to form from fall back of leftover bound envelope. Fall-back times are short (a few $\times 100$ years) and fall-back discs extend between $0.5$ and 5~au (outside the binary orbit), at which point the discs are likely to spread farther on short timescales via viscous torques. These discs  have characteristics in line with those observed. 
\end{abstract}

\section{Introduction}\label{sec:Intro}

Mass transfer in binary systems involving a giant donor and a more compact companion remains one of the most significant "black boxes" in modern astrophysics, with implications on our understanding of any evolved binary population \citep[e.g.,][]{DeMarco2017}. When a giant star fills its Roche lobe, the stability of the resulting mass transfer is dictated by the response of the donor's radius to mass loss compared to the response of the Roche lobe itself (see for example \citet{Soberman1997} and \citet{Hjellming1987}). If the star expands faster than the lobe, the process becomes unstable, leading to a common envelope (CE) phase—a brief, catastrophic episode where the companion is engulfed, leading to dramatic orbital shrinkage or stellar mergers \citep{Ivanova2013}.

Unlike stars with radiative exteriors, which tend to contract or stay nearly the same size upon mass loss, convective donors have an adiabatic response that causes them to expand as they lose their outer layers. 

This expansion creates a feedback loop: as the star grows, it overfills its Roche lobe even further, leading to even higher mass-loss rates. If the Roche lobe cannot expand fast enough to accommodate the growing star—which typically happens if the donor is significantly more massive than its companion—runaway mass transfer ensues, leading to a CE phase.

In this paper we consider a class of post-red giant branch and post-asymptotic giant branch (post-RGB and post-AGB) stars with main sequence companions with semi-major axes of $\sim 0.5-4$~au (periods of $\sim100-2000$~days), eccentricities up to 0.6 \citep{vanWinckel2025} and surrounded by a circumbinary disc \citep{oomen_modelling_2019}.

The orbits of these post-RGB  and post-AGB binaries are typically small enough to imply a recent phase of Roche lobe overflow (RLOF), because their progenitor RGB and AGB stars were larger than the orbital periastron distance. However, a RLOF with a giant should have resulted in a CE interaction with a resulting, much reduced orbital separation, which is not observed for these stars.
 
A possible solution was envisioned by \citet{Soker2015}, who suggested that if during the interaction the companion accretes gas via RLOF and blows jets, it can avoid the CE inspiral and remain in a perpetual phase of non-conservative mass transfer that results in the complete loss of the giant star's envelope. \citet{Kashi2018} later suggested that this "grazing envelope" mechanism can even produce eccentric binaries because of the enhancement of the mass accretion rate near periastron. These scenarios, however, have never been investigated further.

Another way in which the RLOF, mass transfer phase could have been stabilised and have resulted  in the relatively wide separations of post-RGB/AGB binaries was suggested by the discovery that the mean companion mass for these binaries is $1.09\pm0.62$~\ms\ \citep{oomen_modelling_2019}, larger than the typical $\sim$0.5~\ms\ seen in equivalent post-CE, post-AGB binaries such as central stars of PN \citep{Iaconi2019}. These larger companion masses imply larger $q\equiv M_{\rm 2} /M_{\rm 1}$ values at the time of interaction. Larger $q$ values are known to lead to wider post-CE separation in simulations \citep[e.g.][although always below $\sim 30$~\rs]{Passy2012}. It is also known that larger $q$ values may stabilise mass transfer and lengthen the pre-inspiral mass transfer phase. One can then speculate that this may avoid a CE inspiral or lead to a weakened CE.

This said, the exact mass ratio, $M_2/M_1$, above which stability can be assumed is not well known. \citet{Tout1991} found analytically that for a non rotating star, stable mass transfer would take place for a value of $q\equiv M_2/M_1 \gtrsim 1.43$\footnote{\citet{Reichardt2019} erroneously reported  this value to be $q \gtrsim 0.68$ due to a confusion between $M_1/M_2$ and $M_2/M_1$.} (here $M_2$ is the accretor, and $M_1$ is the donor). 

\citet{Ge2015_II} mapped the response of donor stars to rapid mass loss and provided stability criteria for CE precursors. Using adiabatic mass-loss sequences from the zero-age main sequence to the base of the giant branch, they determined that the critical mass ratio for dynamical instability varies strongly with envelope structure. For fully convective donors, this corresponds to $q \approx 1.5 - 1.6$, whereas for radiative envelope donors $q \approx 0.8$, down to $q \approx 0.3 - 0.5$ for more massive radiative donors. Approaching the asymptotic giant branch, the critical mass ratio rises back toward $q \approx 1.3 - 1.7$, making evolved stars far more prone to a CE inspiral. 

Following on from this, \citet{Ge2020_III} showed that superadiabatic convective envelopes can enhance donor expansion and lower the dynamical stability, implying that purely adiabatic stability criteria likely underestimate the tendency of giants to undergo a CE inspiral. However, \citet{Woods2011} found that this instability depends on assuming the superadiabatic surface layer is entirely removed; if mass transfer proceeds on a timescale comparable to the local thermal timescale, the layer can thermally readjust, yielding a more stable response than the adiabatic assumption predicts \citep[see also][]{Temmink2023}. The question of where the stability-instability boundary sits is clearly not yet answered.

In this paper we set out to explore CE interactions between RGB stars and companions with a range of $q$ values, monitoring the phase of RLOF before the CE inspiral, to determine its impact on the final separation and morphology of the interaction. We use the same star as adopted by \citet{Passy2012} and \citet{Reichardt2019}, a 0.88~\ms, 90~\rs, RGB star. The latter work used a compact companion of mass 0.6~\ms\ ($q=0.68$), starting the simulation at the separation that just allows Roche lobe overflow. They found that material leaving the outer Lagrangian points before the inspiral appears to remain mostly bound in a circumbinary disc (see also \citet{Macleod2018b} and \cite{Macleod2020a}). This hinted at the possibility that with a larger companion mass and therefore mass ratio, one may obtain a wider separation as well as the formation of the observed circumbinary discs. In this paper we therefore extend their investigation to higher values of $q$ to increase mass transfer stability, determine whether a circumbinary disc can form, whether a delayed inspiral occurs and whether the final separation will be wider. 

This paper is structured as follows. In Section \ref{Sec:Methods} we outline the simulations used and justify the parameters chosen. In Section \ref{sec:Results} we analyse the binary orbital decay and unbound mass. We then look at the $L_1$ mass transfer rates as a function of both resolution and mass ratio, $q$.  We also investigate the mass lost during the pre-inspiral phase through the $L_2$ and $L_3$ points, and examine the kinematic identity of early ejecta versus that of the ejecta produced during the inspiral, to determine if a disc formed during the early phase can survive the CE. Finally, we look at the possibility of the formation of a post-CE circumbinary disc around these surviving binaries by fall-back of bound gas. In Section~\ref{sec:Discussion} we discuss our findings and explore the implications for the formation of post-RGB binary systems, particularly those documented by \citet{oomen_orbital_2018}, and \citet{kluska_population_2022}. We summarise, conclude, and give details of future investigations in Section~\ref{Sec:Conc}.

\begin{figure*}
    \centering
    \includegraphics[width=\linewidth]{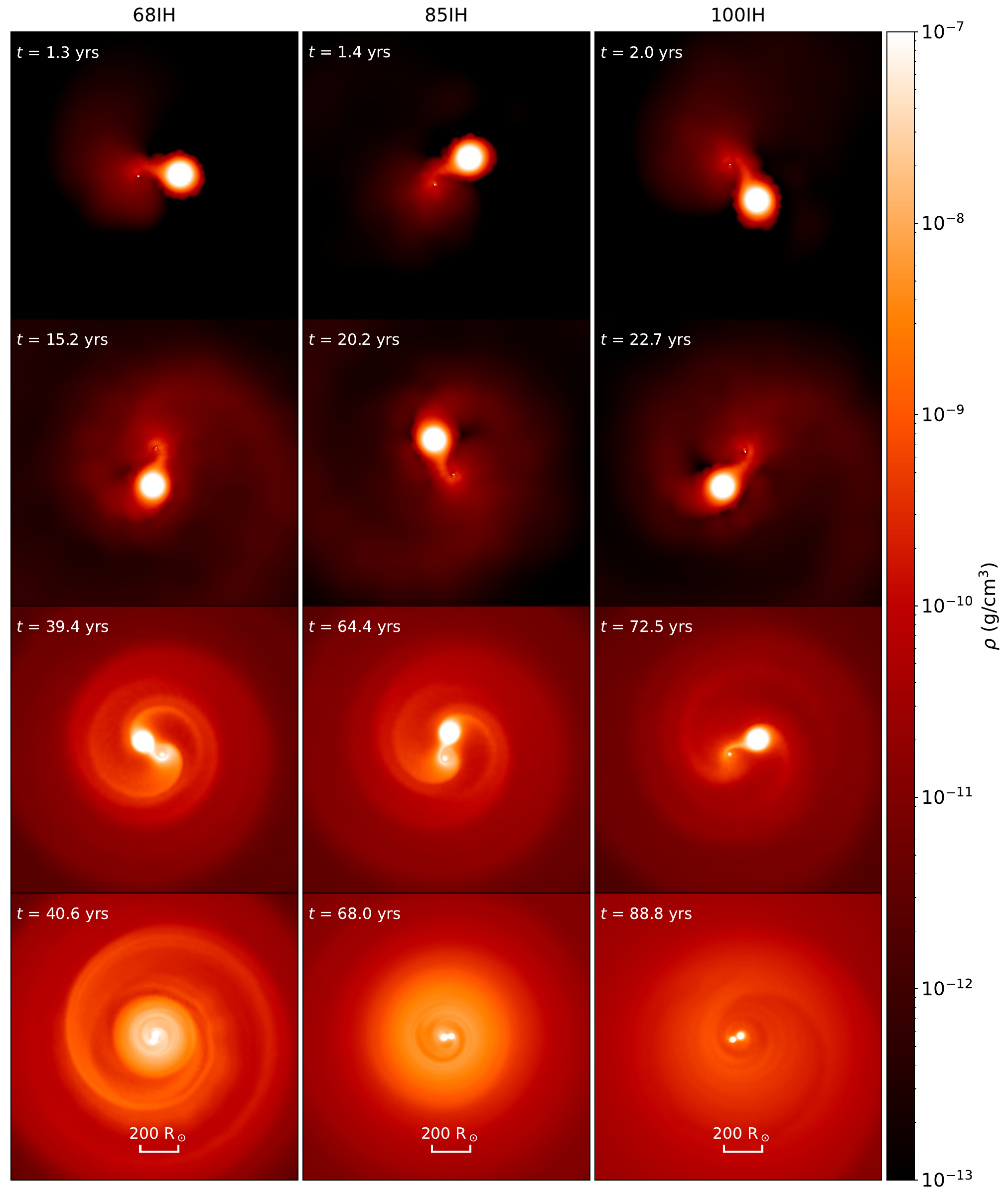}
    \caption{Cross sections of density in the orbital plane of the 68IH (left), 85IH (centre), and 100IH (right) simulations. Each column is a time sequence starting with two moments before the inspiral (top two rows), and ending with the start ($t_i$) and end ($t_f$) of the inspiral (bottom two rows). Each box is approximately 7~au in size.}
    \label{fig:evolution_render}
\end{figure*}

\section{Method}\label{Sec:Methods}

\begin{table*}
    \begin{tabular}{llllllllll}
    \hline
              Sim & $n_{\rm part}$ & $q$ & $M_2$ & $a_0$  & $h_{\rm s}$&$m_{\rm p}$& $t_{\rm end}$&$\alpha_{\rm u}$ &$\alpha_{\rm min}$\\
              Name & ($\times 1000$)&  & (\ms)  & (\rs) & (\rs) &(\ms)& (yr)& &\\
     \hline
              68IL & 76 & 0.68 & 0.60  & 218  & 0.59&$1.2\times 10^{-5}$& 25.3&1&0.1\\
              68IH & 531& 0.68 & 0.60  & 218  & 0.31&$2.0\times 10^{-6}$& 50.5&1&0.1\\
              68MH&531&0.68&0.60&218&0.30&$2.0\times 10^{-6}$&26.2&0.1&0\\
              85IL & 76   &0.85& 0.75 & 230  & 0.59&$1.2\times 10^{-5}$& 40.4&1&0.1\\
              85IH & 531  &0.85& 0.75 & 230  & 0.31&$2.0\times 10^{-6}$& 81.5&1&0.1\\
              85MH & 531& 0.85 & 0.75 & 230 & 0.30&$2.0\times 10^{-6}$& 39.0&0.1&0\\
              100IL & 76   &1.00& 0.88 & 235  & 0.59&$1.2\times 10^{-5}$& 40.4&1&0.1\\
              100IH & 531  &1.00& 0.88 & 235  & 0.31&$2.0\times 10^{-6}$& 105&1&0.1\\
              100MH&531&1.00&0.88&235&0.30&$2.0\times 10^{-6}$&45.5&0.1&0\\
              150IL&76&1.50&1.32&255&0.59&$1.2\times 10^{-5}$&78.4&1&0.1\\
              150IH& 531& 1.50& 1.32& 255& 0.31& $2.0\times 10^{-6}$& 55.6& 1&0.1\\
              150MH& 531&1.50& 1.32 & 255  & 0.30&$2.0\times 10^{-6}$& 148&0.1&0\\
     \hline
              \multicolumn{9}{l}{L and H denote low and high resolution, respectively (see $n_{\rm part}$).} &\\
              \multicolumn{9}{l}{I and M denote the use of an ideal and  tabulated (\textsc{mesa}) EoS, respectively} & \\
    \end{tabular}
    \caption{Simulations' inputs. The second column is the number of SPH particles each simulation uses. The value $q = M_2/M_1$, where $M_2$ is the companion mass and $M_1=0.88$~\ms. The initial separation at Roche lobe contact is given by $a_0$. The value of $h_{\rm s}$ is the  smallest SPH smoothing length at time $t=0$ (noting that the gravitational softening radius for both the primary and secondary core is 3~\rs).  $m_{\rm p}$ is the mass of all SPH particles in the simulation. The time $t_{\rm end}$ is the total simulation physical time. The last two columns are the artificial conductivity and the \textit{minimum} artificial shock viscosity, respectively (the \textit{maximum} artificial shock viscosity is $\alpha_{\rm max}=1$). The differences in these last two columns between the ideal and tabulated EoS simulations are due to the different requirements for stellar stability. }
    \label{tab:simulation_start_summary}
\end{table*}

In order to explore the early pre-CE phase of mass transfer and the characteristics of the post-CE ejecta, we performed 12 simulations using the smoothed particle hydrodynamics (SPH; e.g. \citealt{Monaghan1992,Price2012}) code \textsc{phantom} \citep{Price2018}. 
Across these simulations we have varied $q \equiv M_2/M_1 = 0.68 -1.5$, the resolution (76\,000 and 531\,000 SPH particles), and the simulation's equation of state (EoS; Table~\ref{tab:simulation_start_summary}). Our donor star is the same as that used by \citet{Passy2012}, \citet{Iaconi2017} and \citet{Reichardt2019}: a 1~\ms\ main sequence star that was evolved to the RGB using the 1D code {\sc mesa} \citep{Paxton2010}, till it had a total mass of $M_1 = 0.88$~\ms\ with a core mass $M_{\rm c} = 0.392$~\ms\ and a radius of approximately 90~\rs. We model both our companion star and the primary's core as point mass particles with a 3~\rs\ softened potential. \rev{The 1D stellar structure was mapped into the 3D computational domain. This stellar structure was shown to be stable in the 3D computational domain by \citet{Passy2012}, \citet{Iaconi2017} and \citet{Reichardt2019} by evolving the star in isolation and showing that it remains in reasonable hydrostatic equilibrium for several dynamical times. Later, \citet{gonzalez-bolivar_common_2022} and \citet{Lau2022} improved the stabilisation method, and this is the one we used for both EoSs, obtaining an even more stable star.}

Our lowest mass ratio is $q = 0.68$, the same as used by \citet{Reichardt2019}.  Additional simulations were carried out up to $q=1.5$. A binary with $q=1.5$ implies a 1.32~\ms\ companion, more massive than the main sequence progenitor of the primary. Purely from an evolutionary point of view, we would have to assume the companion in this simulation to be a massive white dwarf star. Observed post-RGB and post-AGB binaries have main sequence, not white dwarf companions, so this simulation is not quite modeling one of those systems. We therefore consider our $q=1.5$ simulation as a technical experiment to push the ratio to a higher value.

In each simulation, we start our binary with a circular $e=0$ orbit with a semi-major axis separation such that the radius of the star is approximately equal to its Roche lobe radius. In doing so, the process of mass transfer from the donor onto the accretor begins immediately in our simulations. This also ensures each simulation begins at Roche lobe contact to model as much of the pre-inspiral phase of evolution as is computationally feasible.

Table~\ref{tab:simulation_start_summary} lists a summary of  parameters for our 12 simulations, where the names are self-explanatory: the number is the mass ratio $\times 100$, "L" and "H" are low and high resolution, respectively and "M" stands for {\sc mesa} EoS as opposed to ideal gas. For each of the four values of $q$ we carried out a low and a high resolution simulations with an ideal gas EoS, as well as an equivalent high resolution simulation with a \textsc{mesa}, tabulated EoS \citep{Paxton2011, Paxton2013, Paxton2015, Reichardt2020}. This additional EoS simulation is key to providing an upper limit to the amount of gas unbound, due to its inclusion of recombination energy, which delivers additional thermal energy to the envelope. The effects of this EoS have previously been studied by \citet{Reichardt2020}, \citet{gonzalez-bolivar_common_2022} and \citet{Lau2022}.

We adopted default artificial conductivity parameter values (see section 2.2.8 of \citealt{Price2018}) of unity for the ideal gas models, and 0.1 for tabulated EoS simulations (see Table~\ref{tab:simulation_start_summary}). 
The default value is set to ensure accurate treatment of contact discontinuities \citep{Price2008}. The conductivity is second order in phantom, with the effective thermal conduction $\kappa_{\rm cond} \propto \alpha_u h^2 (\nabla \cdot {\bf v})$, where $h$ is the resolution length (c.f., \citealt{Price2018}). It thus vanishes when the velocity divergence is zero and the resolution is high, but we found that the remaining heat transfer could nevertheless lead to a slight undesirable expansion of the star over the timescales of interest. The simple solution was to lower the pre-factor for the {\sc mesa} EoS simulations models to $\alpha_u = 0.1$ \citep[see also][]{gonzalez-bolivar_common_2022}.

\begin{figure*}
\centering
    \includegraphics[width=\textwidth]{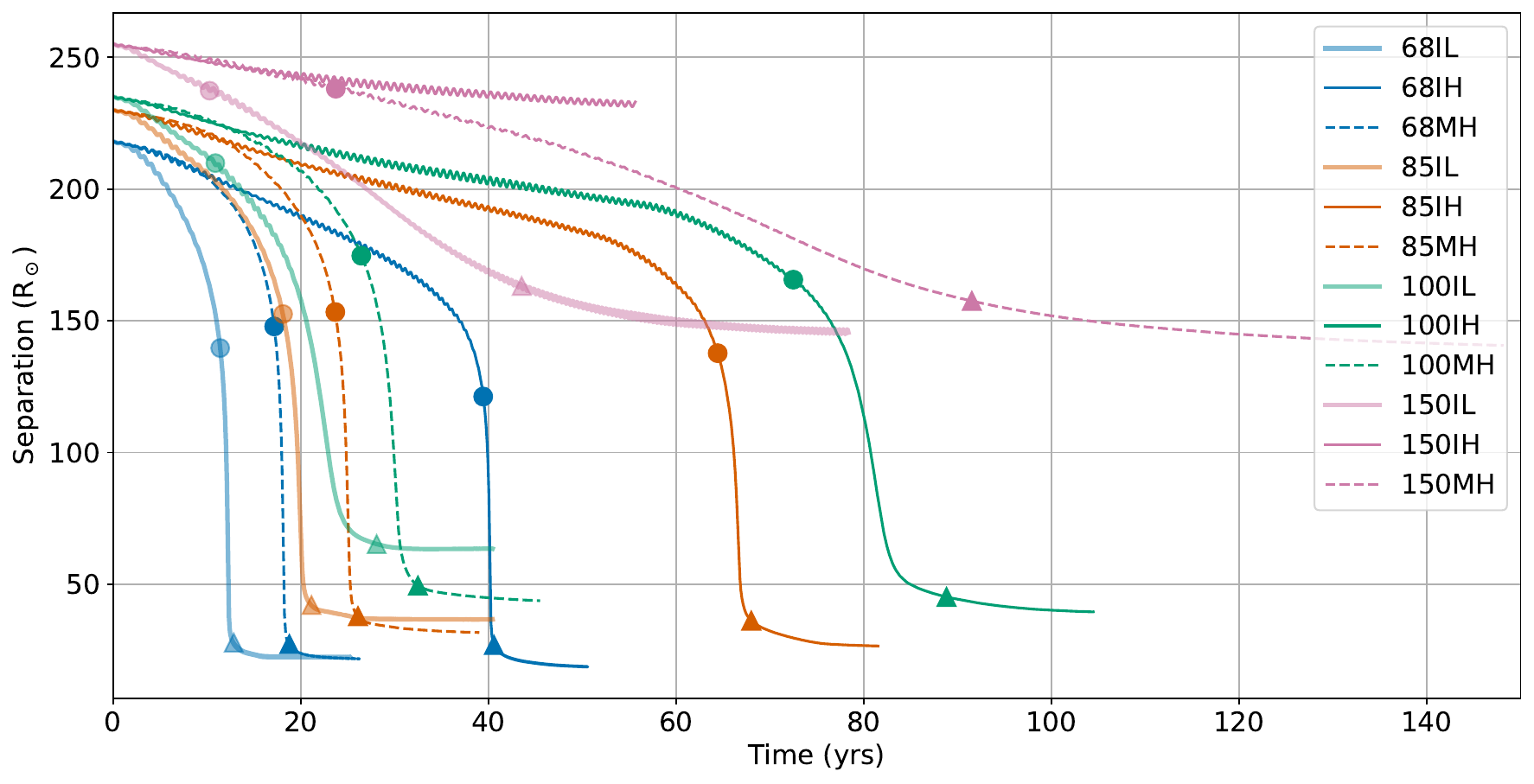}
    \includegraphics[width=\textwidth]{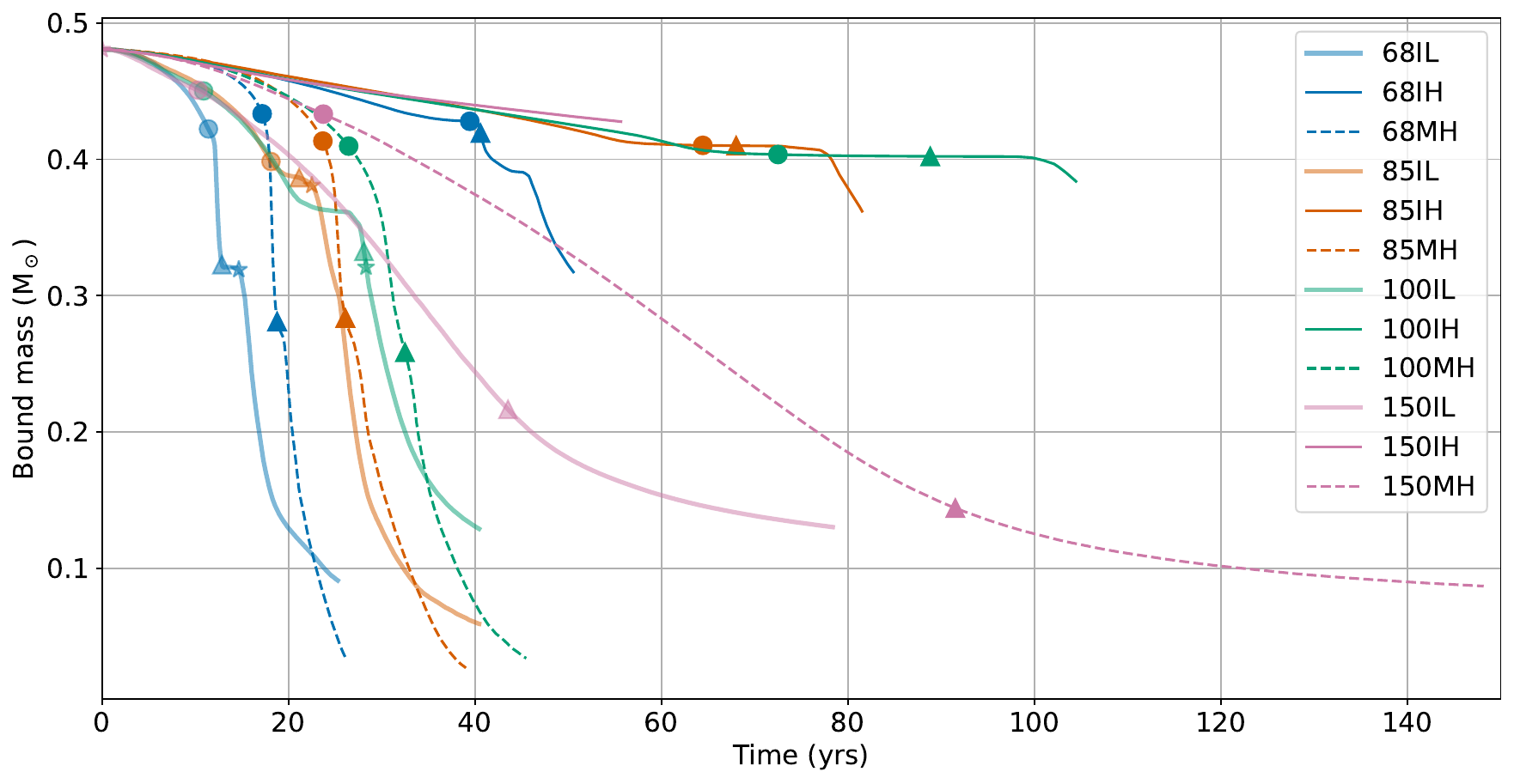}
    \caption{Top panel: binary core separation as a function of time for the twelve simulations (see Table \ref{tab:CE_summary}).  The circles and triangles are the start and end of the inspiral, respectively, as determined by the criterion $|\frac{\dot{a}}{a}| \geq \frac{1}{15} \textrm{max}|\frac{\dot{a}}{a}|$ \citep{Reichardt2019}. Note that this criterion is not adopted for the $q=1.5$ simulations, either due to a very shallow inspiral (150IL and 150MH), or the lack of inspiral (150IH; see text). Extrapolating from the time taken for the 100IL and 100IH simulations to inspiral, the computational cost for continuing the 150IH simulation is currently unfeasible. 
    Bottom panel: the evolution of the bound mass for each simulation. Circles and triangles have the same meaning as in the upper panel, while the stars denote the time at which the resolution-dependent mass unbinding is estimated to start.}
    \label{fig:evo_vs_time}
\end{figure*}
\section{Results}\label{sec:Results}

Here we present the orbital evolution, the characteristics of the ejected mass, and the state of the binary ejecta environment at the conclusion of the CE interaction. In Figure~\ref{fig:evolution_render} we show an overview of the evolution for three of our simulations, 68IH, 85IH, and 100IH, from left to right respectively. We take four snapshots in time, shown vertically, where the first two show progressing RLOF, leading into the start and end of the dynamical inspiral in the latter two panels, respectively.

\subsection{Orbital evolution and unbound mass}
\begin{table*}
\centering
    \begin{tabular}{lllllllllll}
    \hline
         Sim &$a_0$ & $t_\textrm{i}$ & $a_\textrm{i}$ & $t_\textrm{f}$ & $a_\textrm{f}$ & $t_\textrm{steep}$ & $a_\textrm{steep}$ & $\tau_\textrm{steep}$ & $t_{*}$ & $\dot{M}_{\rm {L_1}, i}$\\
         Name & (\rs) & (yr) & (\rs) & (yr) & (\rs) & (yr) & (\rs) & (yr) & (yr) & (\ms/yr) \\
    \hline
            68IL&218 & 11.4& 140& 12.8& 27& 12.2& 72& 0.3& 14.7&$9\times 10^{-4}$\\
            68IH&218 & 39.4& 121& 40.6& 27& 40.1& 67& 0.3&-&$4\times10^{-4}$\\
 68MH& 218& 17.1& 148& 18.8& 27& 18.1& 72& 0.3& -&$4\times10^{-4}$\\
 85IL& 230& 18.1& 152& 21.1& 42& 19.9& 77& 0.7& 22.5&$1\times10^{-3}$\\
            85IH&230 & 64.4& 138& 68.0& 36& 66.6& 69& 0.7& -&$5\times 10^{-4}$\\
            85MH&230 & 23.7& 153& 26.1& 38& 25.0& 78& 0.5& -&$4\times10^{-4}$\\
            100IL&235& 10.9& 209& 28.1& 65& 22.5& 109& 4.2& 28.3&$9\times10^{-4}$\\
            100IH&235& 72.5& 166& 88.8& 45& 81.2& 89& 3.6& -& $7\times10^{-4}$\\
            100MH&235 & 26.4& 175& 32.5& 49& 30.1& 90& 1.3& -&$7\times10^{-4}$\\
            150IL&255& 10.3& 237& 43.5& 163& 56.6& 151& 8.3& -& $2\times10^{-3}$\\
 150IH& 255& -& -& -& 233 \textdagger& -& -& -& -&$9\times10^{-4}$\\
 150MH& 255& 23.7& 238& 91.5& 158& 61.9& 198& 73.6& -&$1\times 10^{-3}$\\
 \hline
 \multicolumn{11}{l}{\textdagger This separation is at the end of the simulation.}
\end{tabular}
    \caption{Summary of parameters relating to the CE inspiral. Here $a_0$ is the initial separation at Roche lobe contact. The beginning and end of the CE inspiral are found using the criterion from \citet{Reichardt2019}, and are denoted  $t_{\rm i}$ and $t_{\rm f}$ respectively, with their associated separations, $a_{\rm i}$, and $a_{\rm f}$. The parameters with the subscript 'steep' refer to the time, separation, and timescale of the point of fastest inspiral in the interaction. The column $t_{*}$ denotes the star in Figure~\ref{fig:evo_vs_time}, approximately the last moment before the resolution-dependent unbinding takes place. $\dot{M}_{\rm L_1, i}$ is the rate of mass transfer onto the companion one year after the start of the simulation.}
    \label{tab:CE_summary}
\end{table*}

Figure \ref{fig:evo_vs_time} shows the orbital evolution and bound mass of our systems as a function of time. We define the beginning and end of the inspiral in the same manner as was done by \citet{Reichardt2019}, that is, the upper and lower bounds of the criterion: $|\frac{\dot{a}}{a}| \geq \frac{1}{15} \textrm{max}|\frac{\dot{a}}{a}|$  (circles and triangles in Figure \ref{fig:evo_vs_time}), where the limiting value, here $1/15$, for the timescale is chosen by reasonable inspection. 
Due to the shallow inspiral of both the 150MH and 150IL simulations this criterion does not capture, even with modifications to the limiting value, satisfactory points that could be considered the start and end of the inspiral. Instead we have chosen an alternative method for these simulations. 

For the other simulations the average rate of orbital decline, $\langle|\frac{\dot a}{a}|\rangle$ is less than 5\% of the simulations $\textrm{max}~|\frac{\dot{a}}{a}|$. For the 150MH and 150IL simulations, however, we estimate it to be approximately 25\%. Given its rate of decline is more linear than the other simulations shown in Figure~\ref{fig:evo_vs_time}, we find the start and end of the inspiral, respectively, as follows: $a_{\rm i} = a_0 - 0.25(a_0-\bar{a})$ and $a_{\rm f} = a_{\rm end} + 0.25(a_0 - \bar{a})$, where $\bar{a}$ is the mean orbital separation over the whole simulation, $a_0$ is the initial separation, and $a_{\rm end}$ is the orbital separation at the end of the simulation. An attempt to determine a universal algorithm that would select meaningful points for all our simulations proved fruitless, so we retain these somewhat arbitrary but quantitative criteria even if for some simulations they do not appear to select reasonable times by visual inspection.  
 
A large value of $q$ leads to a shallower inspiral and  larger final separations (see $\tau_{\rm steep}$ and $a_f$ in Table~\ref{tab:CE_summary}), with the 100IH and 100MH having $a_f \sim 45-50$~\rs. The 150IH simulation shows a decreasing orbital separation, but by 40 years no sign of the dynamical inspiral is present and we stopped this simulation due to excessive computational times. The 150MH simulation, aided by a slightly expanded stellar structure has a gentle inspiral ($a_f\sim 150$~\rs) and plateau, even if the final separation is still declining at up to approximately 0.1~\rs/year by the time we stop the simulation. This also true of the 150IL simulation.

In the bottom panel of Figure~\ref{fig:evo_vs_time} we show the evolution of the bound mass for each simulation (defined as mechanical energy $< 0$). The circles and triangles represent the start and end of the CE inspiral as explained above, while the star symbol in the IL simulations represents the beginning of the resolution dependent unbinding, discussed below. 

Variable amounts of pre-CE mass transfer occur before the rapid inspiral. This phase of early mass transfer typically increases in duration with increasing resolution, and with an increasing value of $q=M_2/M_1$ ($t_{\rm i}$ in Table~\ref{tab:CE_summary}). Note this is not the case for the 100IL, and 150IL simulations due to their significantly smoother inspirals, for which $t_{\rm i}$ by our criteria begins to have difficulty. The duration of this early phase is not converged with resolution, and is generally agreed to be substantially longer in reality \citep{Iaconi2017,Reichardt2019,gonzalez-bolivar_common_2022}. Simulations carried out with the tabulated \textsc{mesa} EoS also exhibit a shorter duration for this early phase of mass transfer compared to that of their ideal EoS counterparts, due to a greater degree of expansion of the star surface layers resulting from a reduced level of surface stability \citep{gonzalez-bolivar_common_2022}.

\begin{figure*}    \centering
    \includegraphics[width=0.495\linewidth]{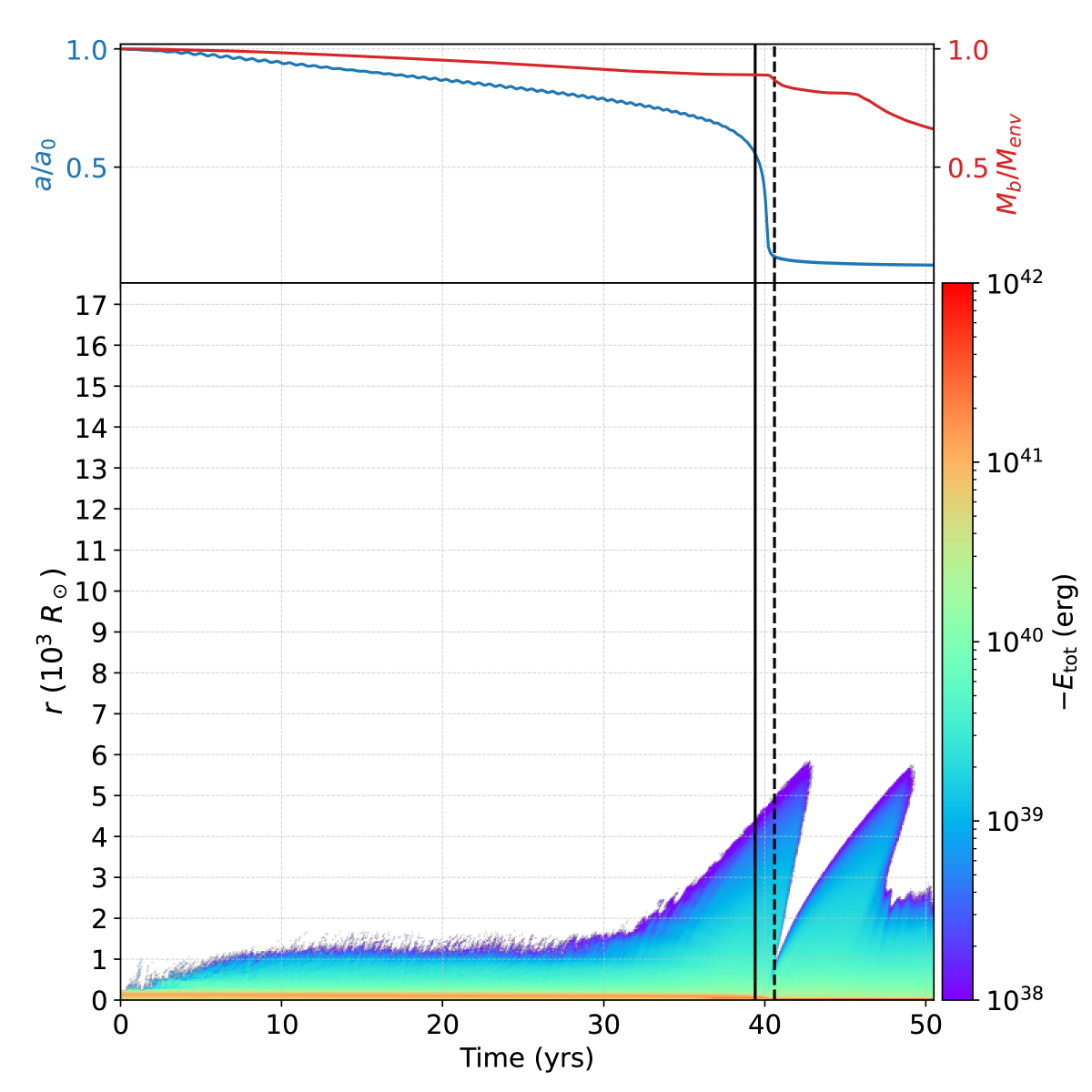}
    \includegraphics[width=0.495\linewidth]{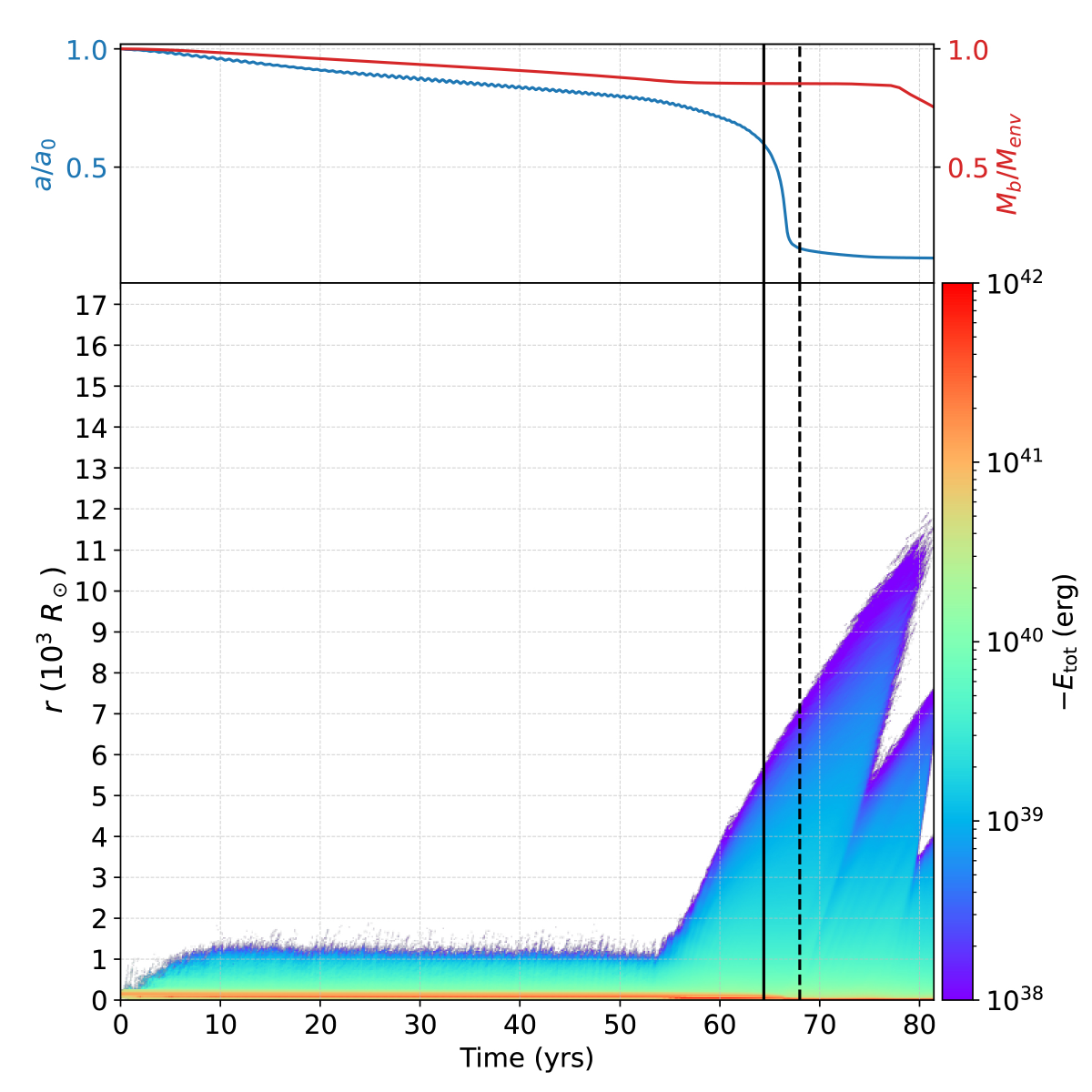}
    \includegraphics[width=0.495\linewidth]{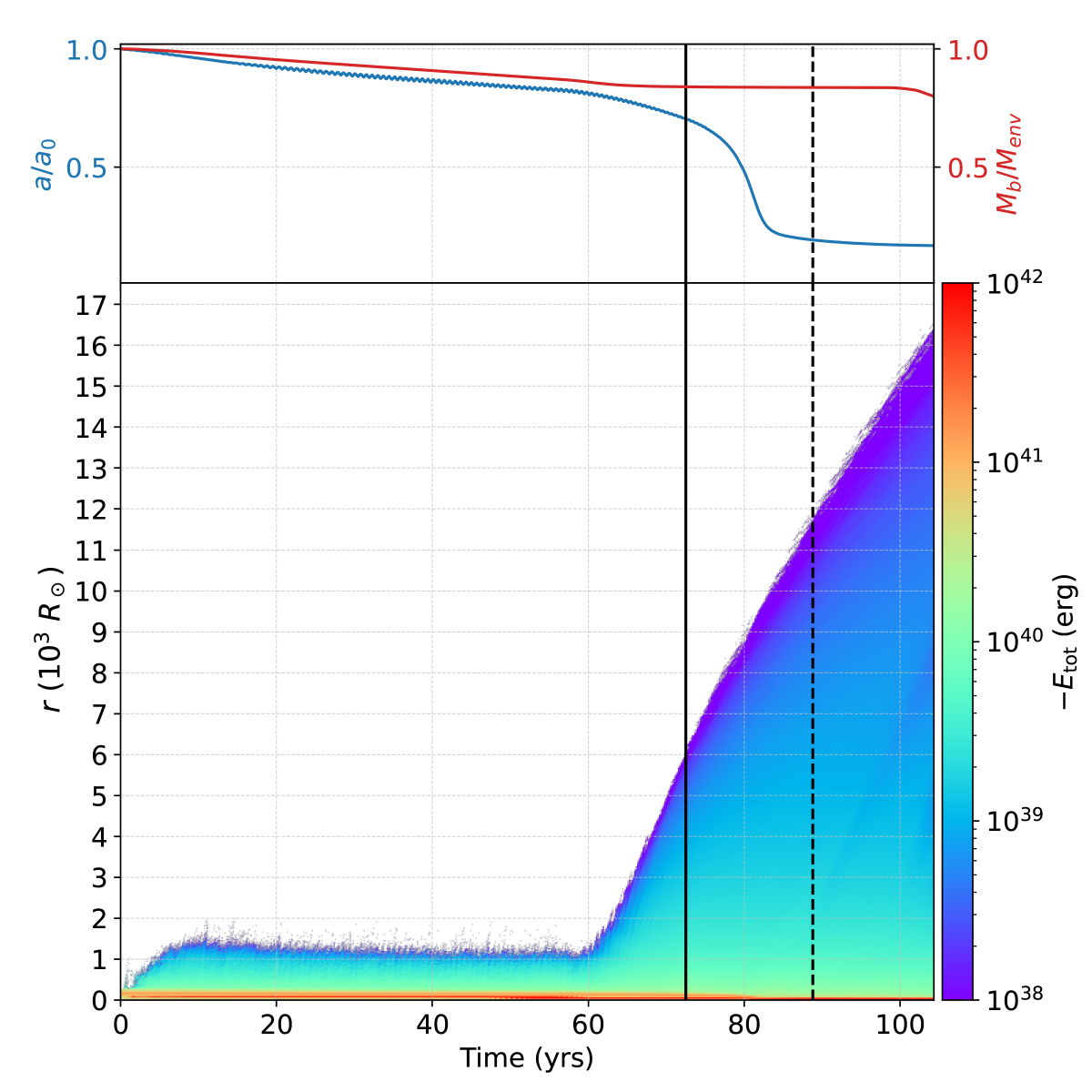}
    \includegraphics[width=0.495\linewidth]{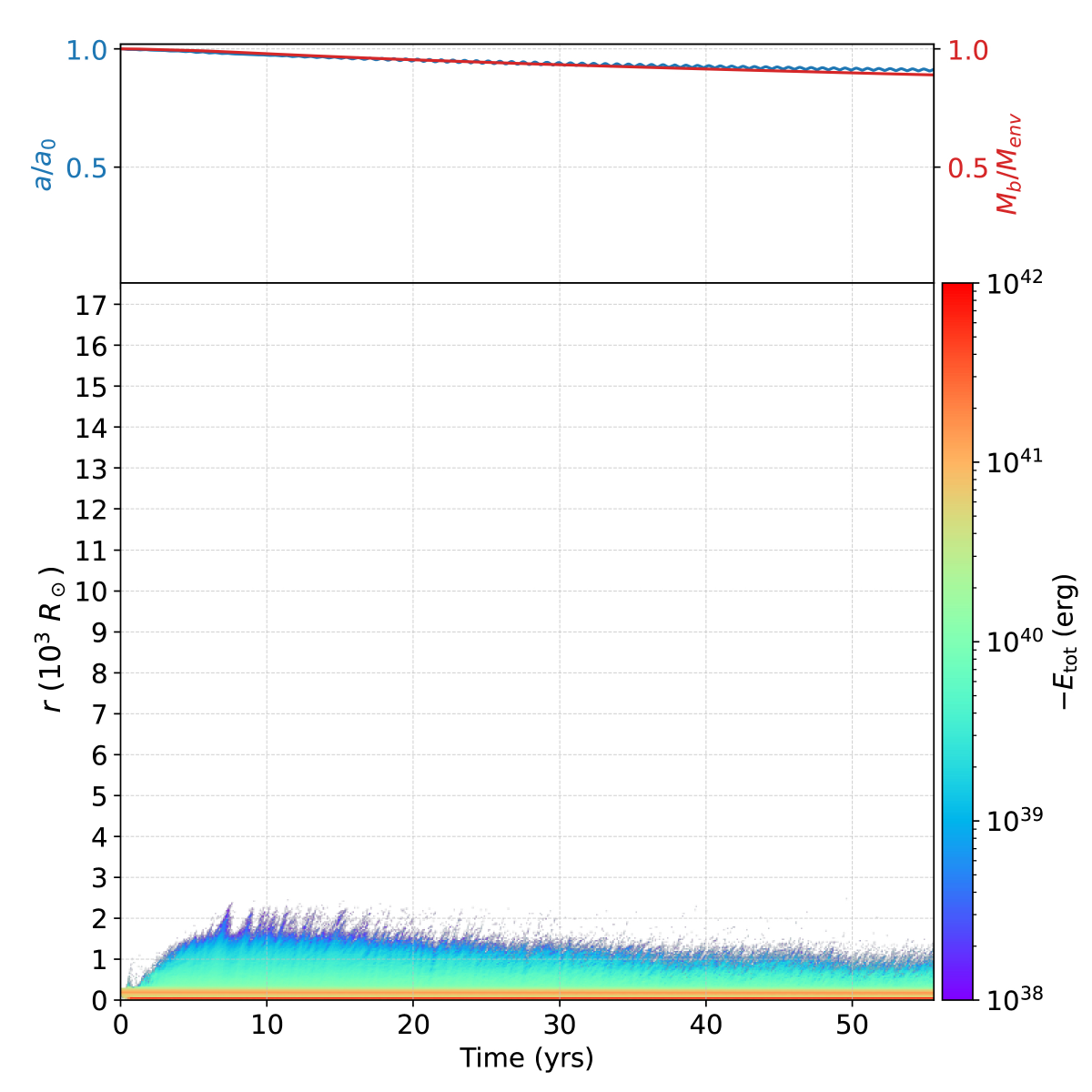}
    \caption{Distribution of bound mass ($K+U<0$) throughout  simulations 68IH (top left), 85IH (top right), 100IH (bottom left), and 150IH (bottom right). The pixels are binned at approximately 10 days in width, and 5~\rs in height, where we calculate the average energy of the gas within that radial bin, at that time step. Top panel: normalised orbital separation (blue) and the bound envelope (red). The vertical lines spanning the two plots denote, from left to right, the start (solid) and end (dashed) of the inspiral. These lines correspond respectively to the circle and triangle in Figure~\ref{fig:evo_vs_time}.    
    }
    \label{fig:LH_bound_inspiral}
\end{figure*}

\begin{figure*}    \centering
    \includegraphics[width=0.495\linewidth]{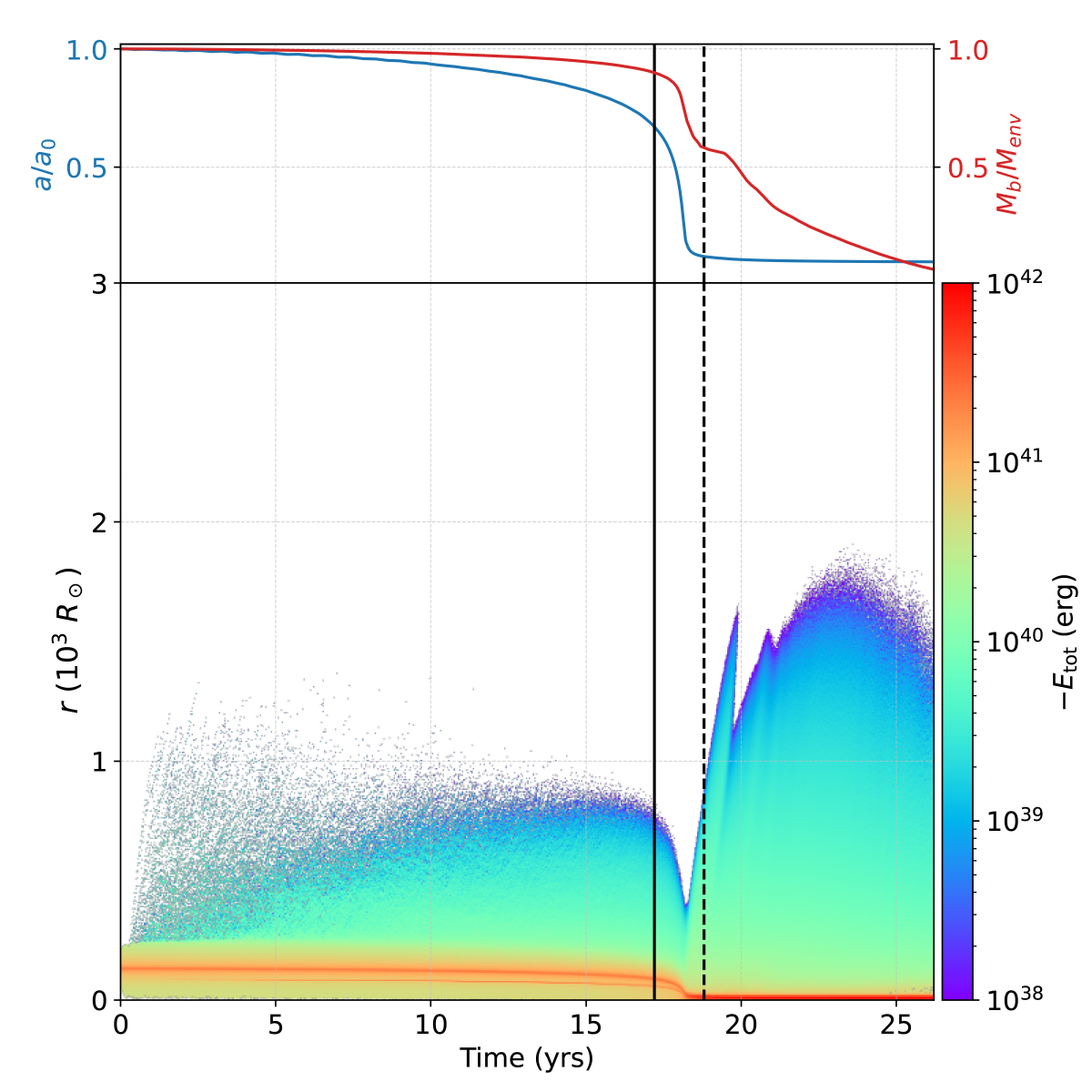} \includegraphics[width=0.495\linewidth]{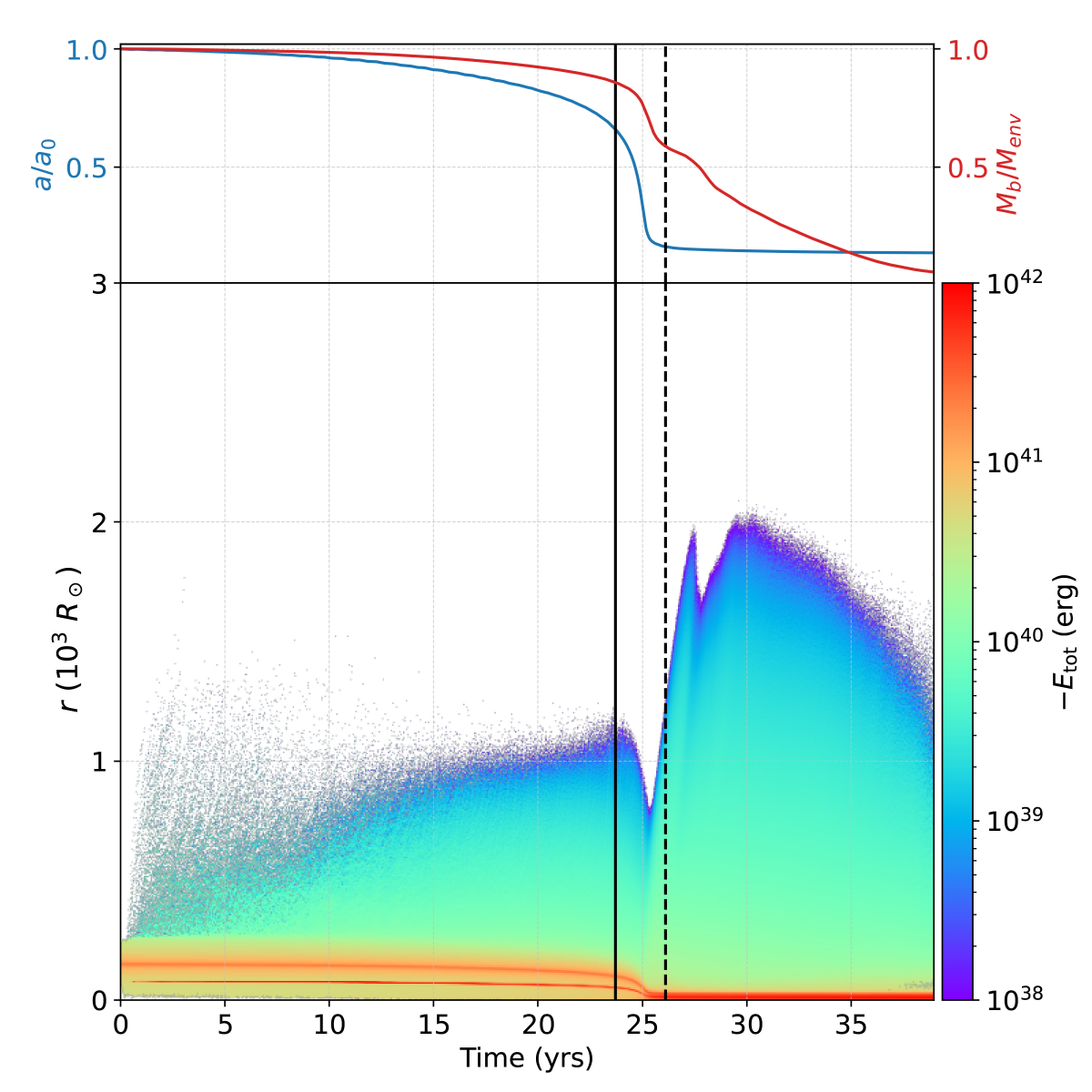}
    \includegraphics[width=0.495\linewidth]{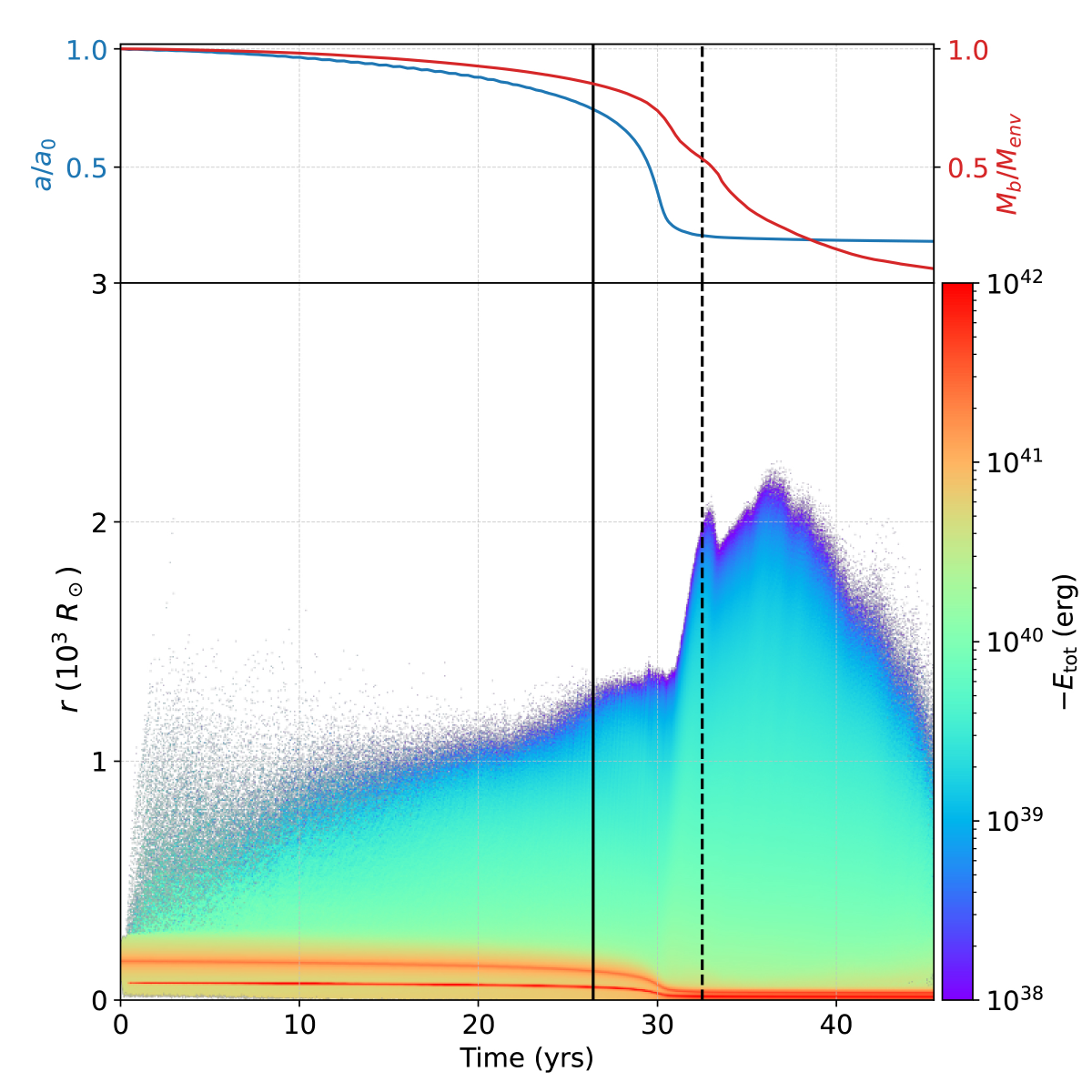} 
    \includegraphics[width=0.495\linewidth]{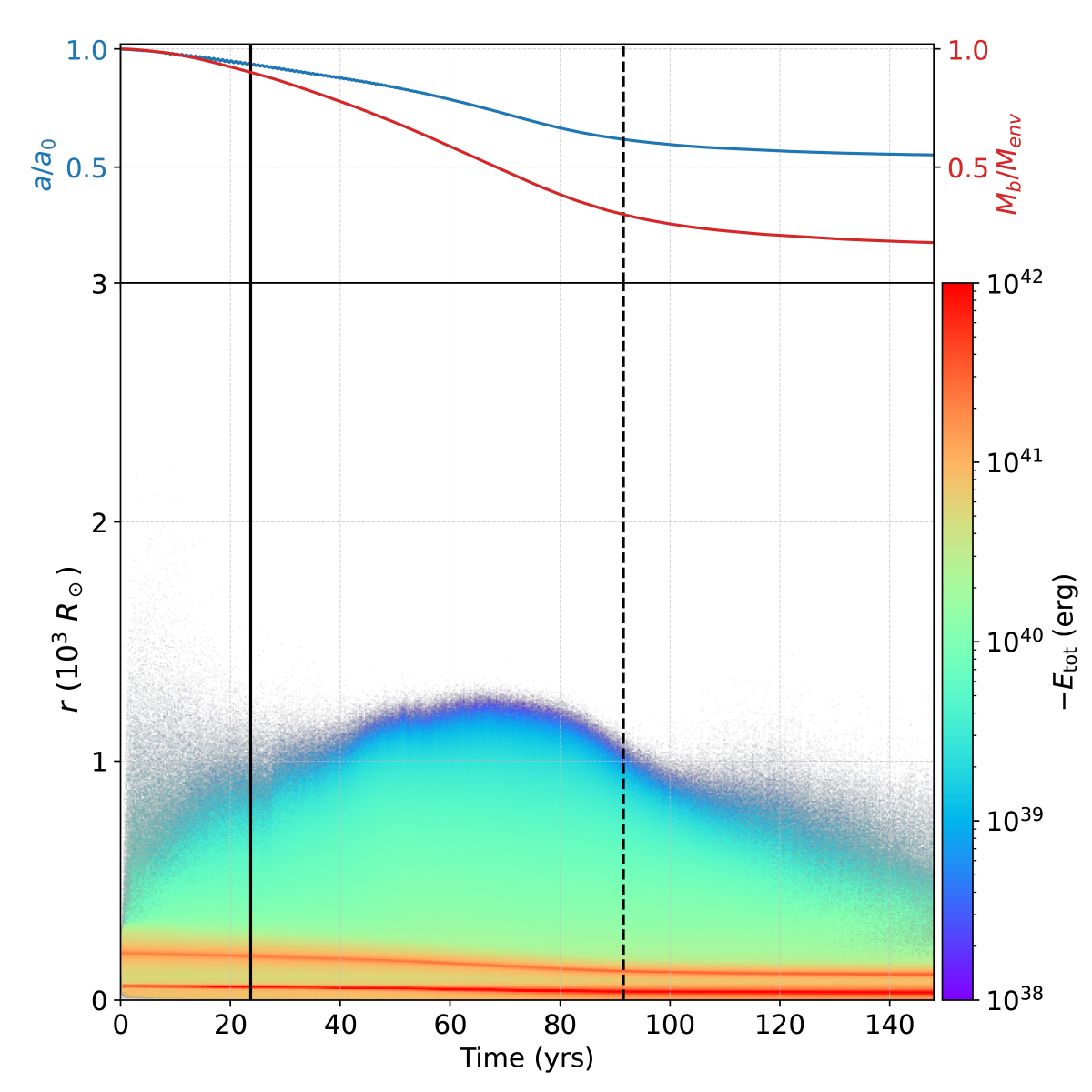} 
    \caption{As in Figure~\ref{fig:LH_bound_inspiral}, but for the 68MH, the 85MH, the 100MH and the 150MH simulations.}
    \label{fig:MH_bound_inspiral}
\end{figure*}

\begin{figure*}
    \centering
    \includegraphics[width=0.7\linewidth]{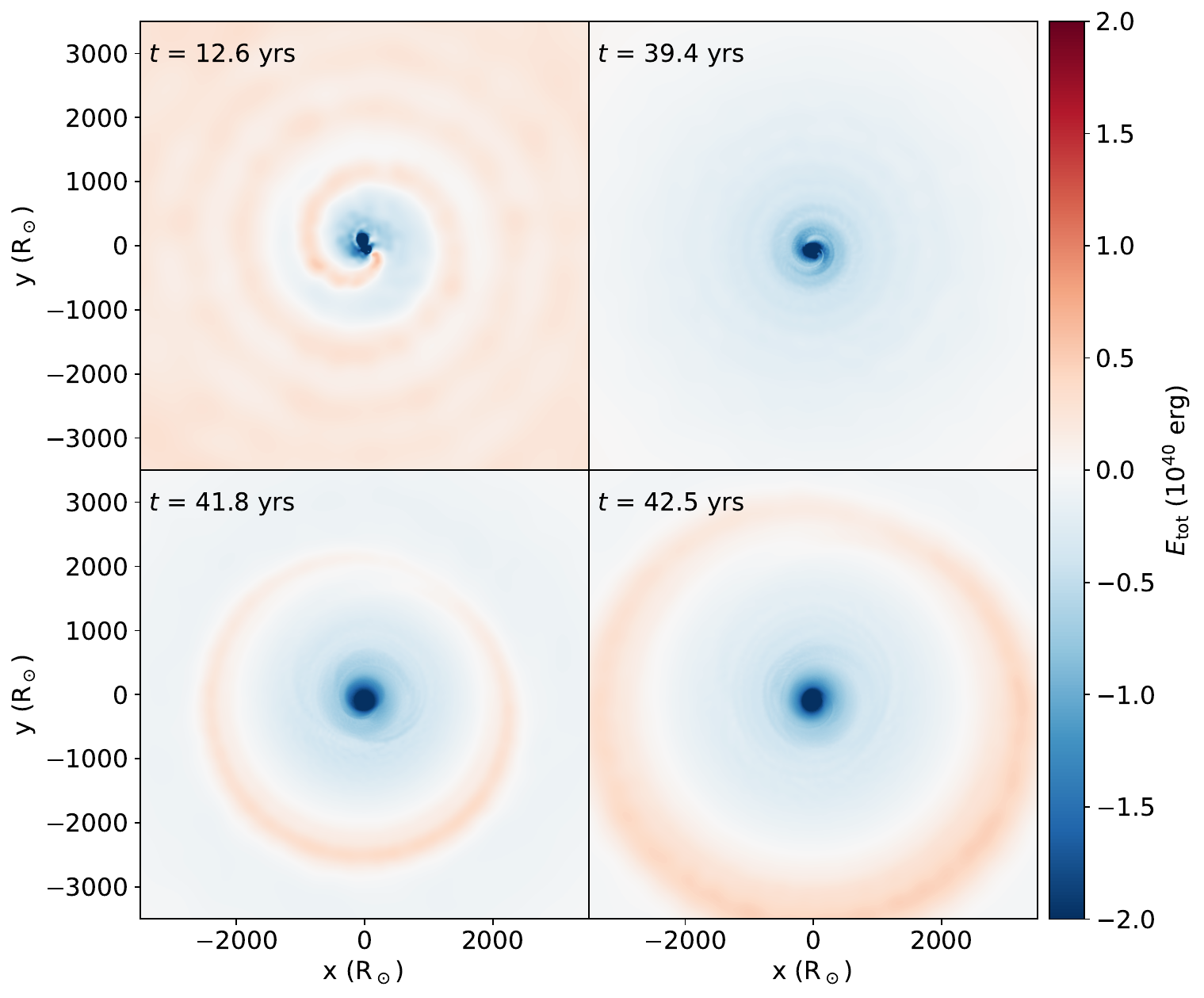}
        \includegraphics[width=0.7\linewidth]{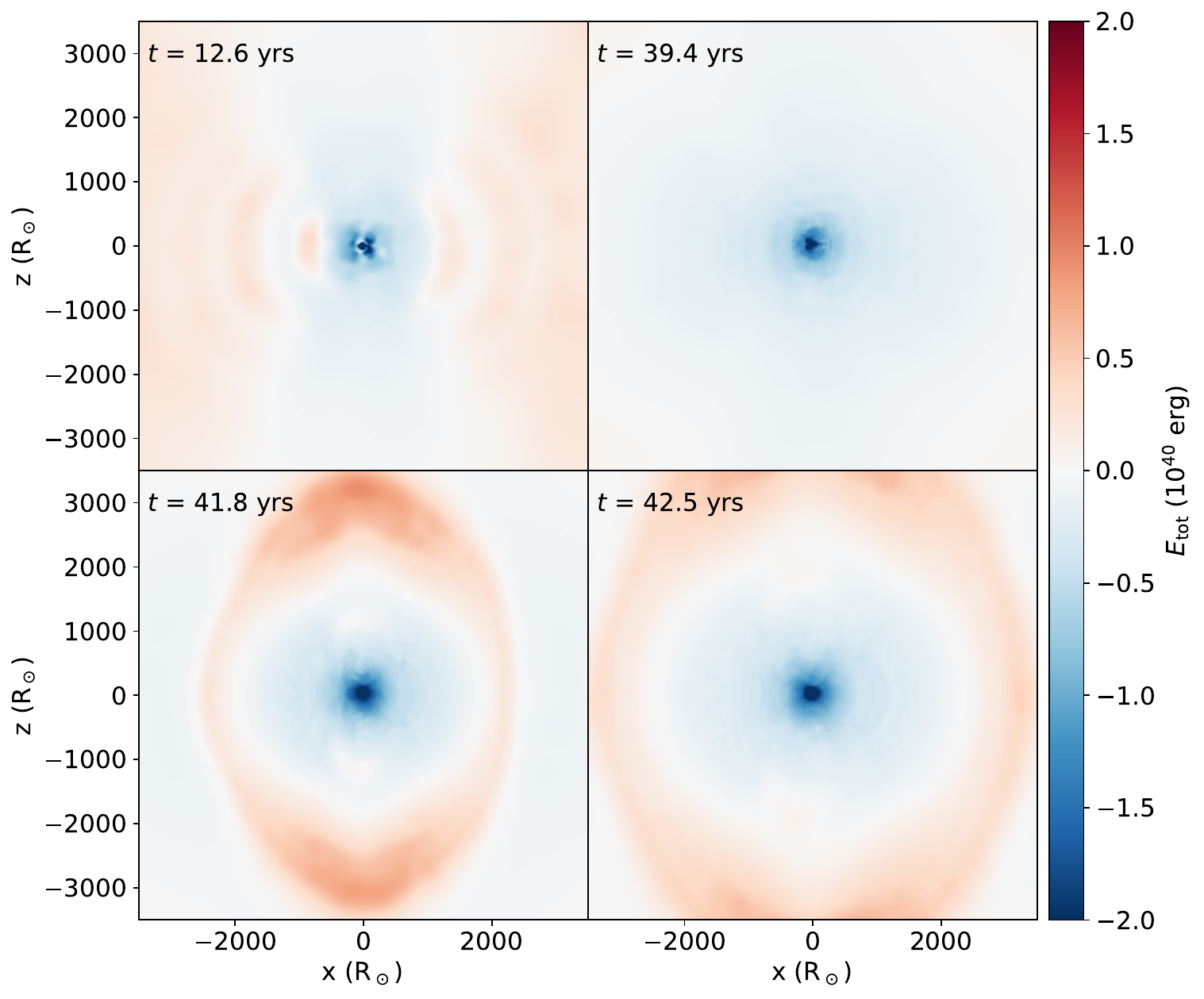}
    \caption{Slices of energy ($E_{\rm tot}$, calculated as in Figures~\ref{fig:LH_bound_inspiral} and \ref{fig:MH_bound_inspiral} but for both positive and negative energies) in the x-y plane (top) and the x-z plane (bottom) for the 68IH simulation. The selected times reflect the early mass transfer period (top left), the start of the inspiral (top right), whereas the bottom two panels depict the unbinding that occurs shortly after the inspiral concludes (as seen after the dashed line in Figure~\ref{fig:LH_bound_inspiral}).
    }
    \label{fig:EN_xy_xz_68H_evo}
\end{figure*}

Higher resolution simulations unbind less mass overall than their lower resolution counterparts (as systematically observed across codes; e.g. \citealt{gonzalez-bolivar_common_2022}). As expected due to the inclusion of recombination energy, the tabulated EoS simulations unbind most of the envelope. This behaviour has been reported before  \citep[e.g.,][]{Reichardt2019,gonzalez-bolivar_common_2022}. 
The additional unbinding seen immediately after inspiral's conclusion completely disappears for high $q$ simulations (present in 68IH but not in 85IH and 100IH), presumably because the envelope is expanded by the deposition of orbital energy but not unbound, which we will continue to investigate shortly. 

The star symbols in Figure~\ref{fig:evo_vs_time} denote the beginning of unbinding due to the density in proximity to the core dropping sufficiently that the SPH particle smoothing lengths become larger than their distance to the core point mass. This leads to additional unbinding already described by \citet{gonzalez-bolivar_common_2022}. This phenomenon is particularly pronounced in simulations with low resolution and an ideal EoS, clear in all the IL simulations (except 150IL) as a pronounced decrease in the bound mass after the star symbol in Figure~\ref{fig:evo_vs_time} . At high resolution this does not occur (see \ref{sec:appen1} for further details).

In Figures~\ref{fig:LH_bound_inspiral} and \ref{fig:MH_bound_inspiral} we plot the bound mass distribution as a function of time for the 68IH, 85IH, 100IH, and 150IH, and 68MH, 85MH, 100MH and 150MH simulations, respectively.  The vertical black lines correspond to the circle and triangle in Figure~\ref{fig:evo_vs_time}, the beginning and end of the inspiral. The total energy of each SPH particle was computed as the sum of its kinetic, core–gas potential, and gas–gas potential energies (making up the mechanical energy), adding the values for all particles, $k$, in each bin, $i$, as follows:

\begin{equation}
    \langle E_{\rm tot}(r_i)\rangle = \frac{1}{N_i}\sum_{k\in i}\left[\frac{1}{2}m_{\rm p}v^2_{k}+U_{k}^{\rm core-gas}+U_{k}^{\rm gas-gas}\right],
\end{equation}
\noindent where $N_i$ is the number of particles per bin, and $m_{\rm p}$ is the mass of each SPH particle. The particles were binned radially into 5~\rs\ intervals, and the mean energy of the particles in each bin was rendered as colour. Note that we do not include thermal energy and only use the more stringent mechanical criterion. This is because the inclusion of thermal energy assumes that the entire thermal energy payload of the stellar envelope will be transformed into bulk kinetic energy. However, as has been shown in other work, this is not necessarily the case \citep[e.g][]{Staff2016,Iaconi2017}.

Only bound material is plotted ($K+U<0$) where we note there is very little bound material with energies larger than the minimum plotted energy of $-10^{38}$~erg. Because each SPH particle has a constant mass, this plotted average energy per particle is equivalent to an average specific energy, since the division by mass introduces only a scaling factor.

An initial expansion of the stellar structure upon the start of RLOF also sees material at the surface becoming loosely bound. In the ideal gas EoS simulations the inspiral delivers orbital energy at the base of the envelope, which causes a wave of unbound material with a time delay that depends on the simulation (the trough between two peaks). Very little material is unbound in these high resolution simulations because the unbound mass at the base of the envelope collides with bound gas above it.

The features in Figures~\ref{fig:LH_bound_inspiral} and \ref{fig:MH_bound_inspiral} show just how complex the dynamics of the unbinding is.  The deep troughs in the upper panels of Figure~\ref{fig:LH_bound_inspiral} are unbinding events, not too dissimilar to those displayed by the low resolution simulations (\ref{sec:appen1}), showing that mass unbinding is not converged property but that, for an ideal gas EoS, less mass is unbound at higher resolution. 
For the tabulated EoS simulations, that are only carried out at higher resolution (Figure~\ref{fig:MH_bound_inspiral}), the unbinding events are less pronounced as unbinding here happens more homogeneously due to the inclusion of recombination energy. For clarity, we note that the high potential energy seen as the horizontal red lines in Figures~\ref{fig:LH_bound_inspiral}  and \ref{fig:MH_bound_inspiral} represent the gas in close proximity of the core particles.

In Figure~\ref{fig:EN_xy_xz_68H_evo} we plot mechanical energies as slices in both the x-y plane (top) and x-z plane (bottom) for the 68IH simulation across four different time steps. These time steps are chosen to show the early, pre-inspiral phase of mass transfer, the start of the inspiral, and then two more time steps shortly after the conclusion of the inspiral. These latter two time steps show the unbinding 'wave' of material that is ejected from the region of the cores as the inspiral ends, which is similarly seen in the top right of Figure~\ref{fig:LH_bound_inspiral} as the spike of unbound material from the base of the envelope between the two peaks of bound material. 

This wave of unbinding travels outwards from the cores and crosses the previously bound material at higher radii, but at the same time leaving bound material in its wake. Due to the primarily equatorial ejection throughout the pre-inspiral mass transfer phase, this later ejecta is also partially obstructed in the x-y plane, favouring polar ejection, as seen in the x-z slices of Figure~\ref{fig:EN_xy_xz_68H_evo}. This unbound outflow through the poles was similarly seen in \citet{gonzalez-bolivar_common_2022}.

\begin{figure*}
    \includegraphics[width=0.75\textwidth]{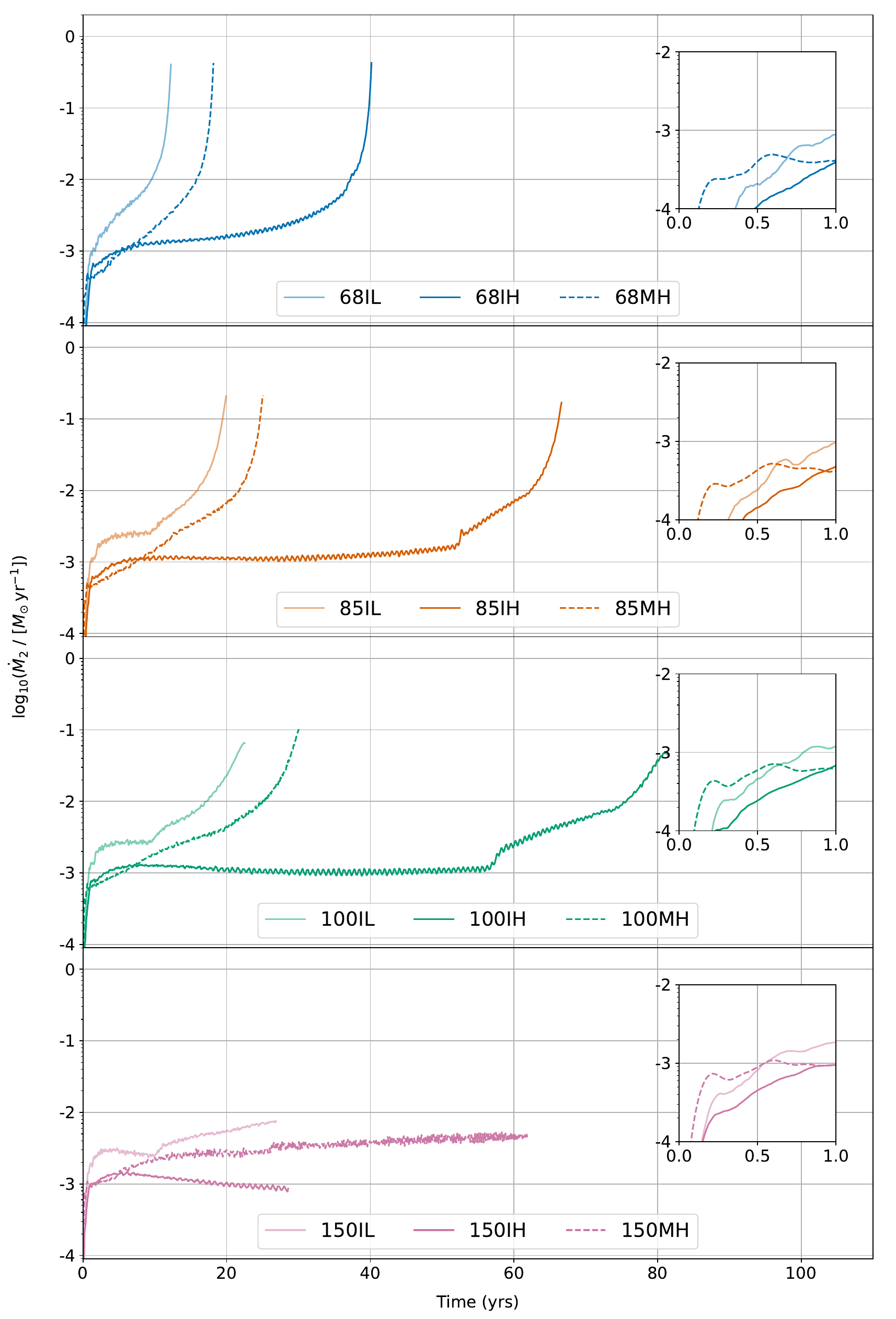}
    \caption{Numerically-derived $L_1$ mass transfer rates as a function of time for each simulation, where the low, high and tabulated EoS simulations, are the dotted, solid, and dashed lines, respectively, in each panel. The calculation is then stopped at the point of steepest inspiral ($t_{\rm steep}$ in Table~\ref{tab:CE_summary}). Inserts zoom in on the first year of mass transfer.}
    \label{fig:Mdot_vs_t}
\end{figure*}

\subsection{The Mass Transfer Rate Through \texorpdfstring{$L_1$}{}}

Figure~\ref{fig:Mdot_vs_t} shows the mass transfer rate through the inner Lagrange point, $L_1$. It begins at the start of the simulation, much earlier than the start of the inspiral and is calculated by counting the number of SPH particles that have passed from the Roche lobe of the donor into that of the accretor between each timestep. For this calculation the mass of the donor and accretor are the sum of their respective sink particle mass, plus any SPH gas particles within their respective Roche lobes. Note that mass transfer may not be conservative as mass may flow through $L_2$ and $L_3$. We end this calculation at the moment of steepest inspiral. 

The initial mass transfer rate ($t \lesssim 1$~yr) appears erratic as it begins with a few particles crossing $L_1$, giving a minimum measurable mass transfer rate \citep{Reichardt2019}. We estimate that we can only calculate a reliable mass transfer rate for $t>$1~year (Table~\ref{tab:CE_summary}). The mass transfer rate at 1~yr for all simulations is in the range $\sim 0.5-1 \times 10^{-3}$~\ms~yr$^{-1}$. The value of the mass transfer rate at the moment of steepest inspiral decreases by approximately an order of magnitude as the value of $q$ increases, from 0.7~\ms~yr$^{-1}$ for $q=0.68$ down to 0.08~\ms~yr$^{-1}$ for the $q=1$ simulations.

The behavior of the $q=1.5$ simulations is distinctive in that the mass transfer rate remains relatively low with values of a few $\times 10^{-3}$~\ms~yr$^{-1}$.  In simulations where $q>1$, mass is transferred from the least to the more massive binary component.  In a conservative evolution, this leads to a widening of the orbit. The simulations 150IL/IH/MH show a modest inspiral likely caused by \rev{a combination of physical mechanisms, including mass loss through $L_2$ and $L_3$,  and tides}.

\citet{Reichardt2019} found (their figure 7) that the analytical  mass transfer rate prescription by \citet{PacSien1972} agreed with their derived values, but this conclusion critically depended on their estimated stellar radius. From 3D SPH simulations, it is difficult to calculate stellar radii accurately, and even more so when the CE interaction brakes the donor' spherical symmetry. It may be meaningless to carry out a comparison with analytical theory in this case but for more details of this problem see \ref{sec:appendixB}.

\subsection{Pre-CE Mass Loss Through \texorpdfstring{$L_2$ and $L_3$}{Mass Loss} }
\label{ssec:pre-ce-mass-loss}

\begin{figure*}
    \centering
    \includegraphics[width=\textwidth]{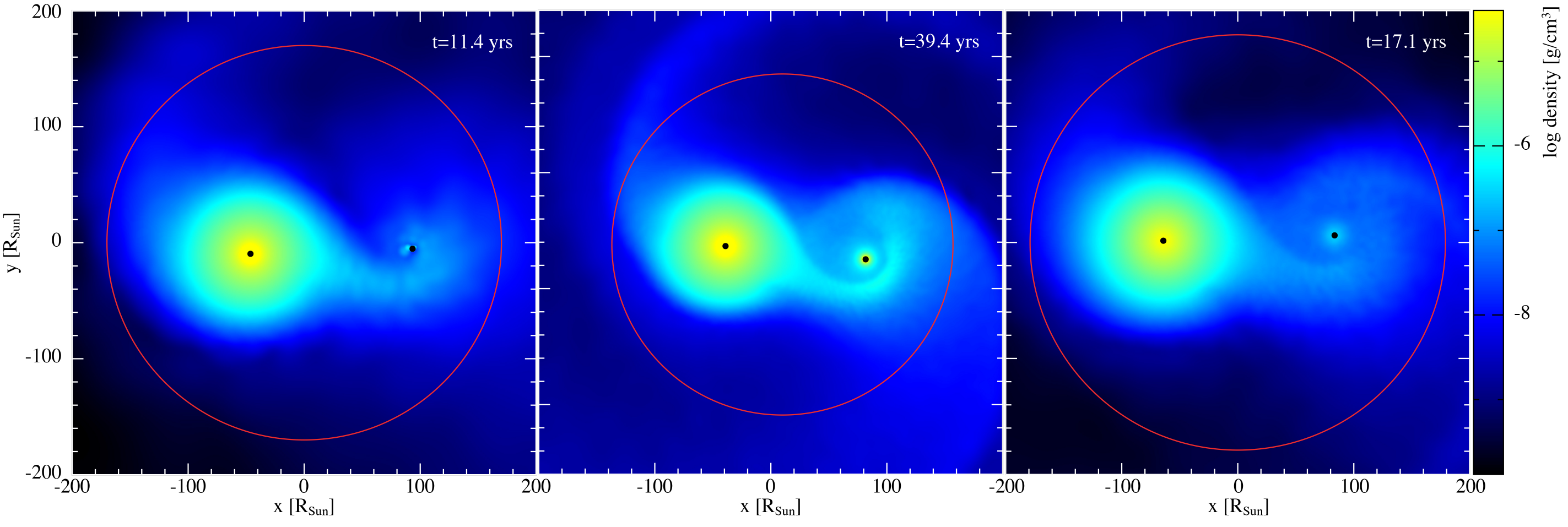}
    \caption{Density slices in the x-y plane of simulations 68IL (left), 68IH (middle), and 68MH (right) at the beginning of the dynamical inspiral (circles in Figure~\ref{fig:evo_vs_time}). The red circle is centered on the centre of mass and passes through $L_2$. Significant mass ejection from behind the accretor ($L_2$, right)  is  accompanied by a slightly less pronounced ejection from behind the donor ($L_3$, left). At high mass ratios such as those used in this work, the difference between $L_2$ and $L_3$ is small. Plot was generated using \textsc{Splash} \citep{Price2007}.}
    \label{fig:L2slice}
\end{figure*}

Mass loss through $L_2$ and $L_3$ can be responsible for the formation of a circumbinary disc and for the reduction of the orbital separation seen during the dynamical inspiral. To estimate the total mass loss from the binary before the onset of the inspiral, we first determine the distance $D_{L_2}$ between the primary star and $L_2$ using the formulation of \citet{Misra2020}:
\begin{equation}
    \frac{D_{L_2}}{R_{\rm L}} = 
    \begin{cases}
     3.334 q^{0.514}e^{-0.052q} + 1.308 & \text{for $q < 1$},\\
     -0.040q^{0.866}e^{-0.040q} + 1.883 & \text{for $q \geq 1$},\\
    \end{cases}    
    \label{eq:L2eq}
\end{equation}
\noindent where $R_{\rm L}$ is the Roche lobe radius of the donor given by \citet{Eggleton1983} formula
\begin{equation}\label{RLapprox}
    {\frac{R_{\textrm{L}}}{a}} = \frac{0.49q^{2/3}}{0.6q^{2/3} + \textrm{ln}(1+q^{1/3})}.
\end{equation}

We then select all SPH particles farther than $L_2$ from the centre of mass ($r_{\rm L_2}$) 
before the time of inspiral, $t_i$, as denoted by the circles in Figure~\ref{fig:evo_vs_time}.  
This pre-inspiral mass loss occurs primarily through $L_2$ (as also seen for varying $q$ values by \citet{Sherbak2025}), but will also increasingly include material ejected through $L_3$ as we approach the onset of the dynamical inspiral. As such, using $r_{L_2}$ captures ejected material through both $L_2$ and $L_3$, providing an upper limit to the total mass lost from the binary. Circles with radius $r_{\rm L_2}$ are shown on density slices in Figure~\ref{fig:L2slice}.

It is possible that the onset of $L_3$ mass outflow may also exert a torque on the binary that results in an additional drain of angular momentum, displayed as a small increase in eccentricity traced by the wiggles on the separation curve present in all simulations (particularly obvious at high resolution in Figure~\ref{fig:evo_vs_time}) prior to inspiral. 

\begin{table*}
    \centering
    \begin{tabular}{llllllllllll}
    \hline
    (1) & (2) & (3) & (4) & (5) & (6) & (7) & (8) & (9) & (10) & (11) & (12) \\ 
         Sim. & $r_{\rm L_2}$ & $t_{\rm i}$ & $M_{>L_2, \textrm{i}}$ & $M_{>L_2, \textrm{UB}, \textrm{i}}$ & {$M_{>L_2, \textrm{UB}, \textrm{i}}$} &$t_{\rm f}$&$M_{>L_2, \textrm{UB}, \textrm{f}}$ & {$M_{>L_2, \textrm{UB}, \textrm{f}}$} & $M_{\rm Tot, UB, i}$ & $M_{\rm Tot, UB, f}$ & $M_{\rm Tot, UB, end}$\\
         Name & (\rs) & (yr) & (\% $M_{\rm env}$) & (\% $M_{\rm env}$) &(\% $M_{>L_2, \textrm{i}}$)&(yr) & (\% $M_{\rm env}$)  &(\% $M_{>L_2, \textrm{i}}$) & (\% $M_{\rm env}$) & (\% $M_{\rm env}$) & (\% $M_{\rm env}$) \\
    \hline
    68IL& 170& 11.4& 17& 12& 71& 12.8& 17& 97& 12& 33& 34*\\
    68IH& 147& 39.4& 23& 11& 47& 40.6& 12& 52& 11& 13&  34\\
 68MH& 179& 17.2& 13& 9.8& 74& 18.8& 13& 100& 9.9& 42&93\\
 85IL& 182& 18.1& 22& 17& 78& 21.1& 19& 89& 17& 20&21*\\
    85IH& 164& 64.4& 29& 15& 51& 68.0& 15& 52& 15& 15&  25\\
    85MH& 182& 23.7& 19& 14& 73& 26.1& 19& 100& 14& 41&  94\\
    100IL& 250& 10.9& 6.7& 6.3& 95& 28.1& 6.7& 100& 6.3& 31&  33*\\
    100IH& 198& 72.5&      27&      16&    60& 88.8&   16&  61& 16& 16& 20\\
    100MH& 209& 26.4& 19& 15& 80& 32.5& 19& 100& 15& 46& 93\\
    150IL& 293& 10.3& 6.4& 6.2& 97&43.5& 6.4& 100&   6.2& 55& 73\\
 150IH\textdagger& 288& -& 12& 11& 96& -& 11& 96& 11& -&-\\
 150MH& 294& 23.7& 11& 9.9& 94& 91.5& 11& 100& 9.9& 70&82\\
    \hline
    \multicolumn{12}{l}{\textdagger~ Did not go through a CE inspiral, percentages are instead taken at the end of the simulation.}\\
    \multicolumn{12}{l}{* Percentage is instead taken at the time denoted by the stars in Figure~\ref{fig:evo_vs_time}.}\\
    
    \end{tabular}
    \caption{Data describing mass lost through $L_2$. Columns are as follows: (2): $r_{\rm L_2}$ --- distance of $L_2$ from the centre of mass at the onset of the dynamical inspiral, (3): $t_{\rm i}$; 
    (4): $M_{>L_2, \textrm{i}}$ --- percentage of the envelope mass outside  $r_{\rm L_2}$ at time $t_{\rm i}$; (5) $M_{>L_2, \textrm{UB, i}}$ --- unbound mass outside of $r_{\rm L_2}$ at time $t_{\rm i}$ as a percentage of envelope mass, and of the mass exterior to $L_2$ (6), respectively; (7): $t_{\rm f}$ --- end of the CE inspiral (triangles in Figure~\ref{fig:evo_vs_time});  $M_{>L_2,\textrm{UB, f}}$ --- mass outside $r_{\rm L_2}$ at $t_i$ that is unbound at $t_f$, as a percentage of envelope mass (8), and as a percentage of mass exterior to $L_2$ (9); (10): $M_{\textrm{Tot, UB, i}}$, and (11) $M_{\rm Tot,UB,f}$ --- total unbound mass in the simulation at $t_i$ and $t_f$, respectively (triangles in Figures~\ref{fig:evo_vs_time}); (12) $M_{\rm Tot,UB,*}$ --- total envelope mass unbound at the end of the simulation. $M_{\rm env} = 0.49$~\ms.}
    \label{tab:L2_b_ub_data}
\end{table*}

Table~\ref{tab:L2_b_ub_data} presents a summary of the bound and unbound gas mass for the material ejected through $L_2$ and $L_3$. \rev{Higher resolution, ideal gas EoS (IH vs. IL simulations) increases slightly the amount of mass outside of $L_2$ by the start of the inspiral ($t_i$), as seen in Column 4, increases dramatically the fraction of that mass that is {\it unbound} (three quarters of the mass is unbound vs. half for lower resolution; Column 6), and results in approximately the same mass remaining unbound by the end of the simulations ($t_f$) - meaning no more of the material outside $L_2$ at the start of inspiral is unbound by the end of it (Column 9).  }

\rev{The tabulated EoS (MH vs. IH simulations) has distinctly less mass outside of $L_2$ by $t_i$ than the ideal gas EoS (but this could be a timing issue), the fraction of that mass that is unbound is higher (three quarters vs. half), but by the end of the inspiral the entirety of the mass outside $L_2$ has been unbound - an effect of the recombination energy delivery, as expected. }

\rev{Finally, looking at the dependency on $q$ for the IH and MH simulations only: the mass outside of $L_2$ by $t_i$ slightly increases along the $q=0.68, 0.85, 1.00$ sequence, between 10-20\% and 20-30\%, and the fraction  of that mass that is unbound at $t_i$ increases somewhat between 50-75\% and 60-80\%. We notice, however that these increases are between $q=0.68$ and 0.85, while there is no real change for the $q=1.00$ simulation compared to the $q=0.85$ one. These effects balance out: the more massive $L_2$ outflows for high $q$ values, are overall less bound, making the potential circumbinary disc mass similar. }

The extreme $q = 1.5$ simulations, with $L_2$ appearing now on the outside of the donor (primary) not accretor (secondary) star, seem to exhibit their own unique behaviour. These simulations regardless of EoS or resolution are extremely efficient at unbinding almost all of the 5-10\% of the envelope they do eject. \rev{Even simulation 150IH shows these trends, despite the fact that it has progressed only 23 years in total and has not gone through the inspiral at all.} Beyond unity, simulations ultimately eject less mass  from the binary, and since virtually all of this material is unbound, we find circumbinary disc formation unlikely.

In columns 10, 11, and 12 we present the total amount of envelope mass unbound at the start of the inspiral, the end of the inspiral, and the end of the simulation, respectively. \rev{Again we find that higher resolution unbinds less mass (Column 11) and the tabulated EoS unbinds generally more (Columns 11 and 12). Low resolution simulations display a strong resolution-dependent unbinding pattern, as discussed previously \citep{gonzalez-bolivar_common_2022}. Interestingly for the 150MH simulations the fraction of unbound mass at the end of the simulation is only $\sim$80\%, although it is decreasing. Fundamentally this information indicates that retaining a fraction of disc mass from $L_2$ ejecta is unlikely.}



\subsection{Kinematic Properties of Mass Lost Through \texorpdfstring{$L_2$ and $L_3$}{Kinematic Properties of Ejecta}}

Here we present a short analysis of the angular momentum that is ejected through $L_2$ and $L_3$ before the inspiral. Caution is needed in carrying out comparisons between simulations, due to the arbitrary definition of the time of inspiral start. In particular the definition for the 150IL/IH/MH simulations is substantially different, as is the nature of the inspiral.
\begin{table*}
    \centering
    \begin{tabular}{ccccccccc}
    \hline
    (1) &(2)& (3) &(4)  &(5) &(6) &(7)&(8)&(9)\\
           Sim.&$M_{\rm i}$& $J_{\rm i}$&$\gamma_{\rm loss}$&$\gamma_{ L_2}$&  $M_{\rm UB}$& $J_{\rm UB}$  & $h_{\rm binary}$ &$J_{\rm binary}$\\
           Name&(\% total)& (\% total)& &&(\% total)& 
    (\% total)& $\times 10^{19}$(cm$^2$s$^{-1}$) &$\times 10^{52}$(g cm$^2$s$^{-1}$)\\
    \hline
 68IL& 5.5& 22& 4.0 &4.8&3.9& 17  & 1.3 &3.9\\
 68IH& 7.6& 28& 3.7 &4.3&3.6& 15  & 1.3 &3.9\\
 68MH& 4.3& 18& 4.2 &4.9&3.2& 14  & 1.3 &3.9\\
 85IL& 6.4& 22& 3.4 &4.4&5.0& 18  & 1.5 &4.7\\
 85IH& 8.4& 31& 3.7 &4.1&4.3& 18  & 1.5 &4.7\\
 85MH& 5.7& 22& 3.9 &4.4&4.1& 16  & 1.5 &4.7\\
 100IL& 1.8& 8& 4.4 &5.5&1.7& 8  & 1.5 &5.4\\
 100IH& 7.3& 28& 3.8 &4.7&4.4& 19  & 1.5 &5.4\\
 100MH& 5.1& 18& 3.5 &4.9&4.0& 14  & 1.5 &5.4\\
 150IL& 2.1& 3.9& 1.9 &6.2&2.1& 3.9  & 1.7 &7.5\\
 150IH\textdagger& 2.5& 6.9& 2.8 &6.2&2.4& 6.7  & 1.7 &7.5\\
 150MH& 2.2& 6& 3.1 &6.2&2.3& 7& 1.7 &7.5\\
 \hline
 \multicolumn{9}{l}{\textdagger~ Did not go through a CE inspiral, value is instead taken at the end of the simulation.} \\ 
 \end{tabular}
    \caption{Properties of the $L_2/L_3$ ejecta material outside  $r_{L_2}$ at the start of inspiral. Columns 2 and 6 are taken from columns 4 and 5 of Table~\ref{tab:L2_b_ub_data}, but now show the percentage of the total binary mass (including the core of the primary and companion) outside $L_2$ at the start of the inspiral; columns 3 and 7 are the angular momentum outside $L_2$ at this time, as well as the angular momentum that is outside $L_2$ \textit{and} unbound, as a percentage of total angular momentum of the system. The parameter $\gamma_{\rm loss}$ is defined in \citet{Nelemans2000} and shows the ratio between the specific angular momentum of the material lost from the binary, and the initial specific angular momentum of the binary. \rev{We also calculate $\gamma_{L_2} = h_{L_2}/h_{\rm bin}$, where $h_{L_2}$ is the initial specific angular momentum of $L_2$}. In the final two columns we provide the binary's specific and total angular momenta, respectively. }
    \label{tab:M&J_L2}
\end{table*}


Similar to what was found in Section~\ref{ssec:pre-ce-mass-loss}, each simulation ejects around  $ 4-8$\% of the total binary's mass through $L_2$ by the start of the inspiral (Table~\ref{tab:M&J_L2}, column 2; \rev{where we express percentages of total mass instead of envelope mass as was the case in Table~\ref{tab:L2_b_ub_data}}). Of the mass lost through $L_2$ and $L_3$ , about half is unbound in the IH simulations, and almost all is unbound in the MH simulations, as already discussed in Section~\ref{ssec:pre-ce-mass-loss}. In column 3 of Table~\ref{tab:M&J_L2} we calculate the total angular momentum of the mass that is lost from $L_2$. The IH simulations eject $\sim$30\% of the binary's total angular momentum leading up to the inspiral, with the IL simulations ejecting somewhat less and likely demonstrating a lack of convergence. IL simulations, but not IH, exhibit a trend of decreasing fraction of ejected angular momentum for larger values of $q$. The MH simulations eject a smaller, $\sim 20\%$, with the 150MH showing a much larger value of 55\%, the inverse trend to the $L$ simulations (the 150IH cannot be considered due to the short time of the simulation).  

\rev{A convenient way to express the amount of angular momentum that is lost ($\Delta J$) from a binary due to mass loss ($\Delta M$) is with the parameter $\gamma_{\rm loss}$ \citep{Nelemans2000}, which can be defined as: }

\begin{equation}
    \frac{\Delta J}{J_{\rm bin}} = \gamma_{\rm loss}\frac{\Delta M}{M_{\rm tot}}.
\end{equation}
\rev{The quantity $\gamma_{\rm loss}$ quantifies the impact of mass loss on the evolution of the orbit. Larger values of $\gamma_{\rm loss}$ imply each percentage of mass lost carries with it a higher percentage of the binary's angular momentum, thus eliciting a more significant change in the binary's orbit. }

\rev{In Table~\ref{tab:M&J_L2} (Column 4) we list $\gamma_{\rm loss}$ (calculated with values with subscript "i", hence at the start of in-spiral) to be in the range of 3 to 4 for most of our simulations. This indicates that the escaping material carries away 3-4 times the mean specific orbital angular momentum of the binary, implying relatively efficient removal of angular momentum for a given amount of mass loss. We also see a slight decrease of $\gamma_{\rm loss}$ as mass ratio increases for MH simulations, though the IH simulations have a pretty constant value. This may imply that, for a given amount of mass loss, higher mass ratio binaries remove angular momentum slightly less efficiently, experiencing slower orbital evolution. Our extreme $q=1.50$ simulations tell a complex story. Both ideal gas simulations have considerably smaller $\gamma_{\rm loss}=2-3$.}

\rev{Using binary evolution reconstruction techniques based on three observed double white dwarf systems, \citet{Nelemans2000} determined that for the first of two interactions that generated those systems today, a stable mass transfer phase between a $\sim 2$~\ms\ giant and a similar mass main sequence companion (with no CE inspiral) was characterised by $\gamma_{\rm loss} \sim 1.7$. Larger values of around $\gamma_{\rm loss} \approx 3-7$ (modulated by orbital evolution - a decline from $\sim7$ to $\sim3$ for the initial slow orbital decline, followed by an increase back to $\sim 7$ during the fast inspiral) were instead determined by \citet{MacLeod2018a} for three binary simulations  with $q=0.3$. \citet{Macleod2020a} calculated overall larger values, with simulations with $q=0.03-0.3$ yielding values between $\sim 21$ and $\sim$6 (lower $\gamma_{\rm loss}$ for larger $q$, similar to what we observe). }

\rev{We also calculate $\gamma_{L_2} = h_{L_2}/h_{\rm bin}$, where $h_{L_2}$ is the specific angular momentum of $L_2$ at the start of the simulation when it is at its maximum and $h_{\rm bin}$ is the total specific angular momentum of the binary at time zero (Table~\ref{tab:M&J_L2}, Column 8). We find that in all cases, $\gamma_{\rm loss} <\gamma_{L_2}$, consistent with both \citet{MacLeod2018a} and \citet{Macleod2020a}. We note that $\gamma_{\rm loss}$ need not be constant, as shown by \citet{Macleod2020a}, while we have calculated what is effectively an average value over the slow in-spiral before the CE fast inspiral takes place.}


\begin{figure*}
    \centering
    \includegraphics[width=0.45\linewidth]{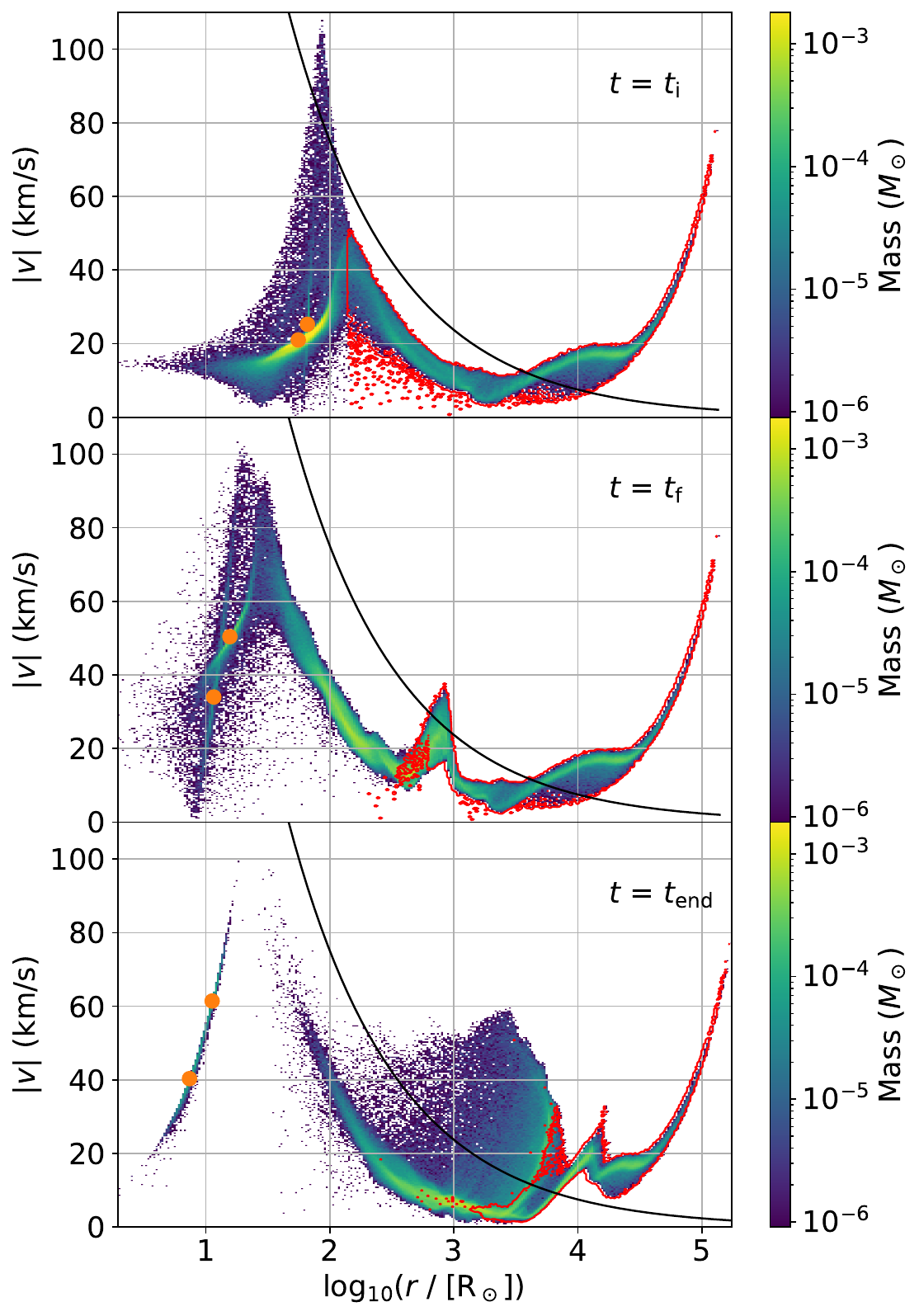} \includegraphics[width=0.45\linewidth]{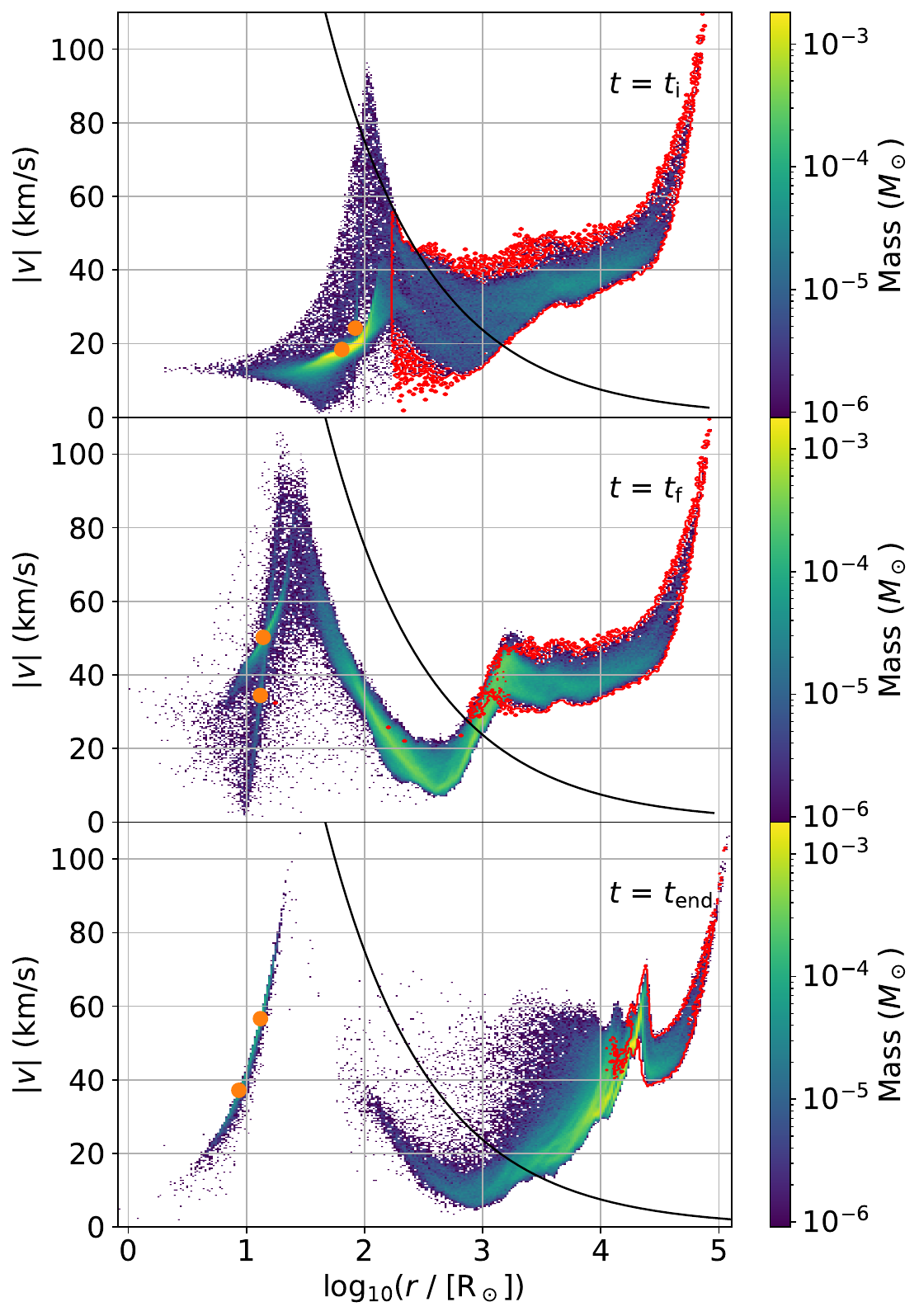}
    \caption{Velocity of each SPH particle as a function of distance from the binary's centre of mass for the 68IH (left) and 68MH (right) simulations. The times shown in chronological order from top to bottom are the start of the inspiral, the end of the inspiral, and the last timestep of the simulation. For simplicity the black line is an approximate escape velocity that assumes the central mass is the primary star {\it and} the companion --- providing an upper limit for bound material. The particles within the red contour are located outside the radius of $L_2$ at the onset of the inspiral as shown is  Figure~\ref{fig:L2slice}. To give an indication of how the material is distributed we construct a 2D histogram of mass with 300 bins in each axis, where the mass shown is the mass per bin. We have also marked the core and companion particles in orange.}
    \label{fig:68_VvsR}
\end{figure*}

Figure~\ref{fig:68_VvsR} shows the velocity of each SPH particle as a function of distance from the centre of mass for the 68IH (left panels) and 68MH (right panels) simulations at the start (top row) and end (middle row) of the inspiral, and at the final timestep of the simulation (bottom row). The particles ejected from $L_2$ and $L_3$ before the onset of the dynamical inspiral are contoured in red and correspond to those outside the red circle in Figure~\ref{fig:L2slice}. The black curve in these plots represents the escape velocity, separating bound from unbound gas.

The material with ($v> v_\mathrm{esc}$) that is unbound at small radii at the time of the inspiral's onset (top row), collides with, and pushes out the material above it. This causes some of the inner ejecta to slow down and possibly become bound once again. In doing so it is also accelerating the outer ejecta, where a peak of unbound gas develops at the boundary between the $L_2$ ejecta and material that follows after, ejected during the inspiral.

Unbound material expands nearly homologously (the $L_2$ outflow is arranged in a straight line in a linear-linear plot). The MH simulations unbind almost all the $L_2/L_3$ ejecta, while the ideal gas EoS simulations do not but the material is swept by the ejecta that comes later, following the inspiral. As such, the material that was ejected through the outer Lagrange points does not have the opportunity to form a disc. We conclude that disc formation via $L_2$ and $L_3$ ejection is unlikely if a CE ejection takes place. Next we investigate whether a fall-back disc is possible.

\subsection{A Fall-back disc}

When the orbit stabilises after the interaction, the fraction of the envelope that remains bound may fall back. Depending on how much envelope falls back onto the binary, and where it falls back to, it may cause a second interaction (as simulated by \citet{Kuruwita2016}) that could result in a further change of the binary orbit. If, however, the material has sufficient angular momentum, it could form a circumbinary disc instead. 

We start by assuming that the ejected, but bound material is in a ballistic orbit, such that both angular momentum and orbital energy are conserved. This material will then reach the pericentre of its orbit within an orbital period. We take the  fall back time for this initial phase to be half of an orbital period, on average:
\begin{equation}
    t_{\rm fb} = \pi\sqrt{\frac{a^3}{GM}},
    \label{eq:t_fb}
\end{equation}
where $M$ is the total central mass (primary core and companion), $G$ is the gravitational constant and $a$ is the semi-major axis of the orbit (derived from each SPH particle velocity vector, distance from the centre of mass and central mass). The fall back radius coincides with the circularisation radius and is given by
\begin{equation}
    R_{\rm c} = \frac{h^2}{GM},
    \label{eq:Rc}
\end{equation}
$h$ is the specific angular momentum.

As gas streams back toward the center of mass and approaches the fall-back radius, collisions between particles generate shocks and dissipation of kinetic energy. The orbit circularises, but angular momentum is conserved. Circularisation timescales are of the order of the orbital period at this new semi-major axis given by $R_{\rm c}$
\begin{equation}
\tau_{\rm circ} \sim  2\pi \sqrt{\frac{R_c^{3}}{GM} }.
\end{equation}

Eventually the system will further evolve through viscous interactions, which will lead to the disc spreading - some material falling into the potential well and accreting to the binary while some material moves outwards. The associated timescale is
\begin{equation}
    \tau_{\rm visc}=\frac{R_{\rm c}^2}{\nu},
    \label{eq:tVisc}
\end{equation}
where $\nu$ is the gas viscosity which can be written as
\begin{equation}
    \nu = \alpha c_{\rm s}^2\sqrt{\frac{R_{\rm c}^3}{GM}},
    \label{eq:nuviscosity}
\end{equation}
\noindent where $c_s$ is the sound speed and $\alpha$ the disc viscosity parameter \citep{Shakura1973}. We now use these estimates of fall-back radius, $R_{\rm c}$ (henceforth $R_{\rm fb}$), and timescale, $t_{\rm fb}$, to approximate the size and distribution of any post-interaction disc that might form. For the mass of this disc, we take the bound mass present at the end of the simulation.

Ideal gas simulations likely underestimate the amount of unbound mass (and overestimate the amount of bound mass), while tabulated EoS simulations would do the opposite since all liberated recombination energy is allowed to do work and none escapes. In this way IH and MH simulations likely bracket the available mass to make a fall-back disc. 

\begin{figure}
    \centering
    \includegraphics[width=\linewidth]{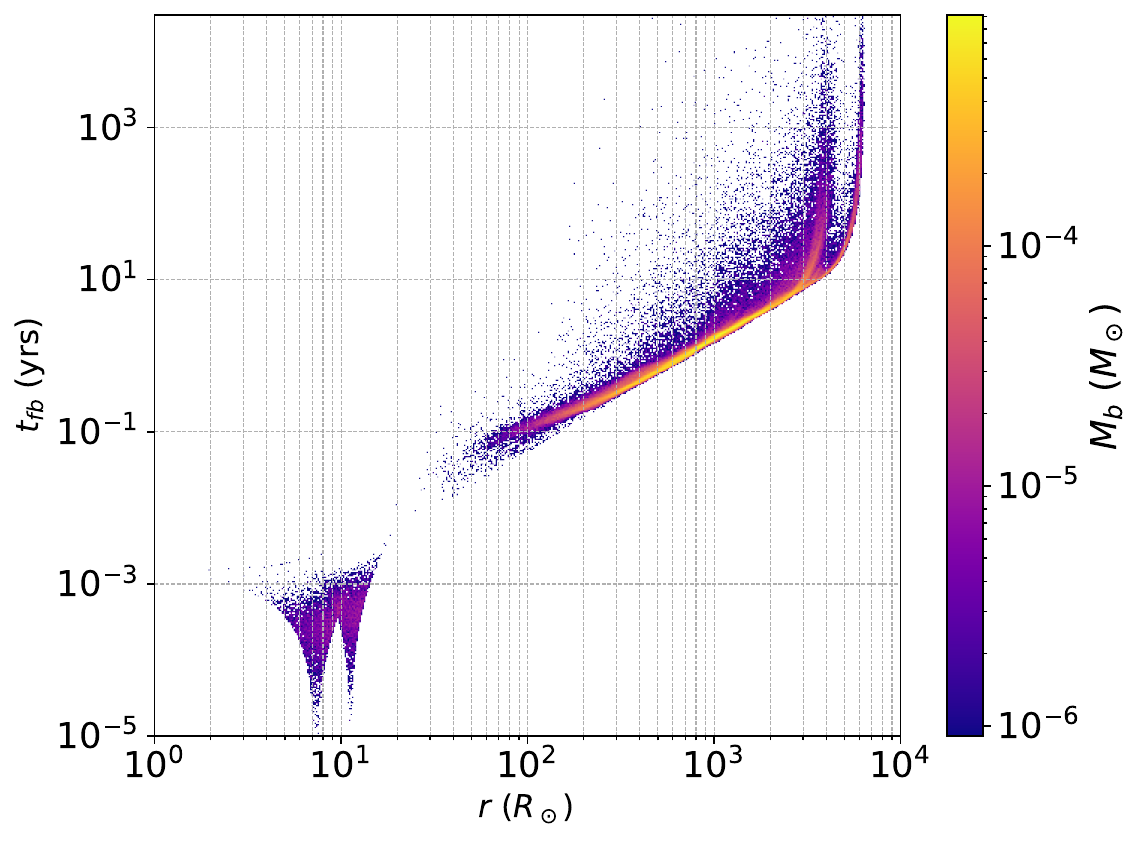}
    \includegraphics[width=\linewidth]{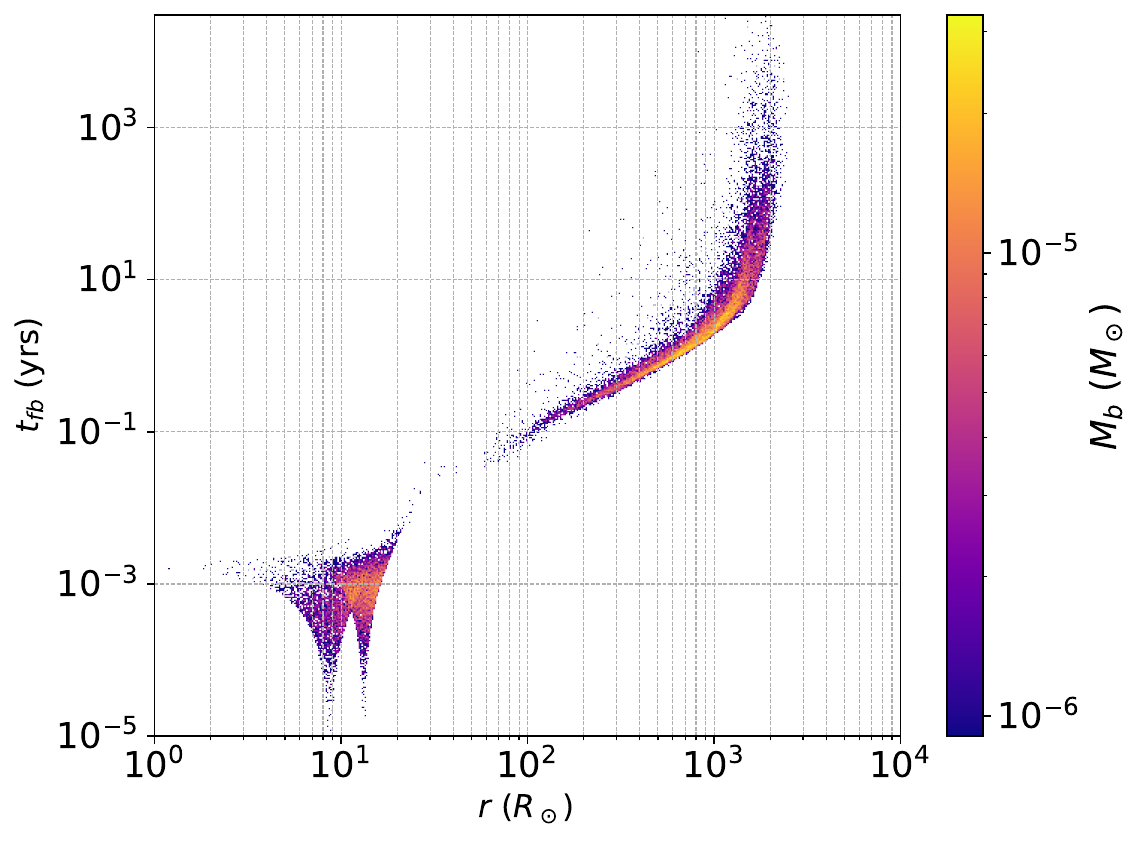}
    
    \caption{Fall-back time as a function of distance from the centre of mass for our 68IH (top), and 68MH (bottom) simulations. The colour bar indicates the amount of mass within each logarithmically sized bin (pixel), to show the distribution of bound ejecta at $t=t_{\rm end}$. To calculate the fall back time we take half the orbital period, where the semi-major axis, $a$, is derived from the orbital energy of the gas particle.
    }
    \label{fig:68_fb_timescale}
\end{figure}

To calculate the fall-back timescale given by Equation~\ref{eq:t_fb}, we take $a$ to be the semi-major axis of the orbit for each SPH particle, calculated from the specific orbital energy, $a = -GM/2E$. Here we calculate $E$ in a similar manner to what was done previously, by summing the specific mechanical energies for the bound material $E = 0.5v_{\rm p}^2 + U_{\rm p, core} + U_{\rm p,p} < 0$, where $v_{\rm p}$ is the velocity of the SPH particle, and $U_{\rm p, core}$ and $U_{\rm p, p}$ are the specific potential energies of the particle due to both the core particles and the other SPH gas particles. In doing so the current motion of each bound SPH particle is considered, as opposed to just its distance from the centre of mass. For our 68IH/MH simulations, we plot $t_{\rm fb}$ as a function of current location from the centre of mass in Figure~\ref{fig:68_fb_timescale}. In this plot we also logarithmically bin the mass of each particle according to these axes, thus giving an indication of how the mass is distributed at this point in time when we might expect material to begin falling back.

As shown for simulations 68IH and 68MH, we find that the mass available to fall back is initially distributed up to a distance of a few $\times 1000$~\rs\ at the end of the simulation. Material with longer fall-back times can be seen at larger radii and corresponds to gas that is moving away from the binary at speeds that are close to escape velocity. Although not shown here, a similar distribution of fall back material is found for all simulations.

\begin{figure}
    \centering
    \includegraphics[width=\linewidth]{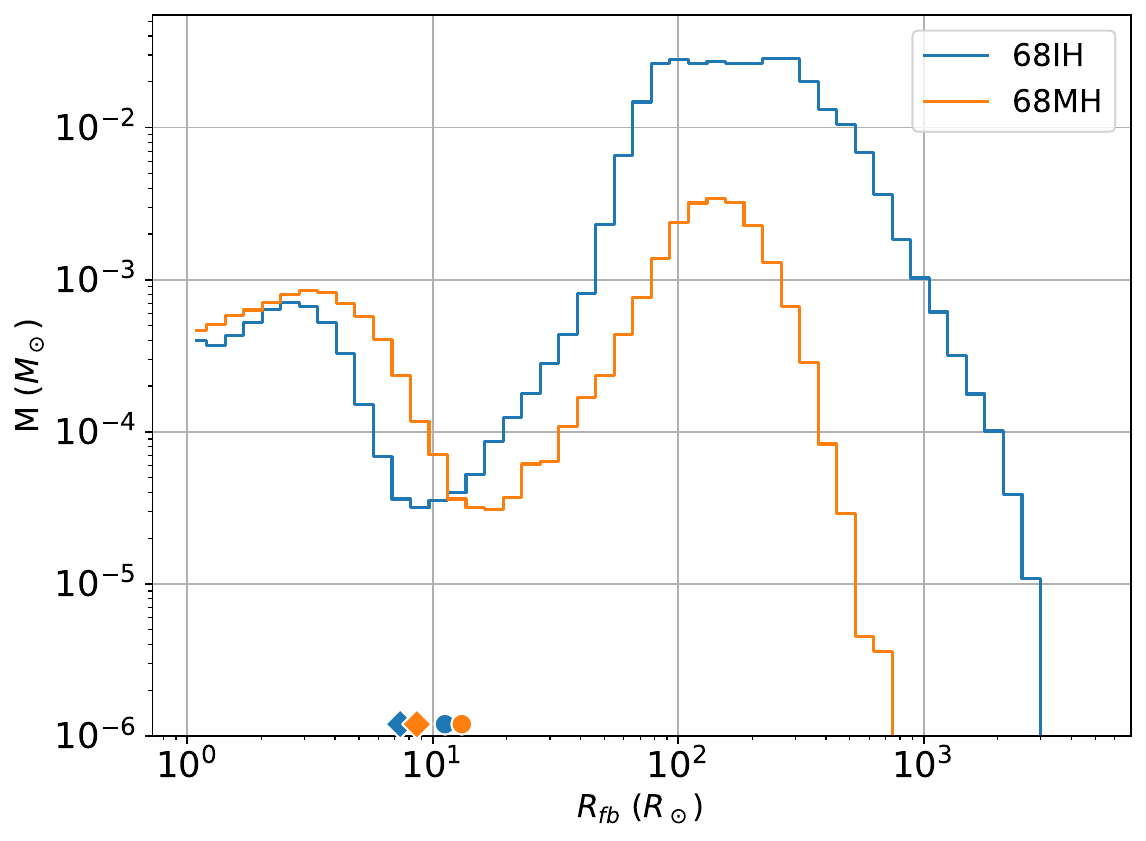}
    \includegraphics[width=\linewidth]{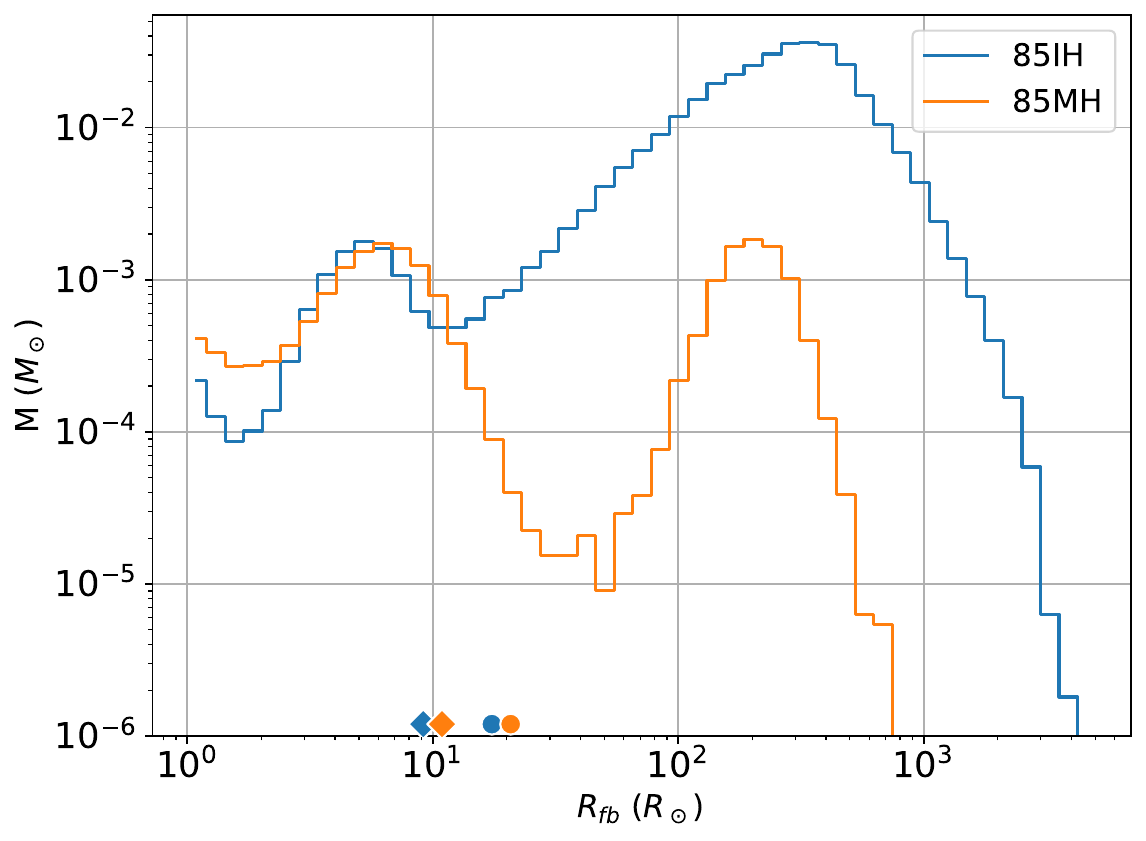}
    \includegraphics[width=\linewidth]{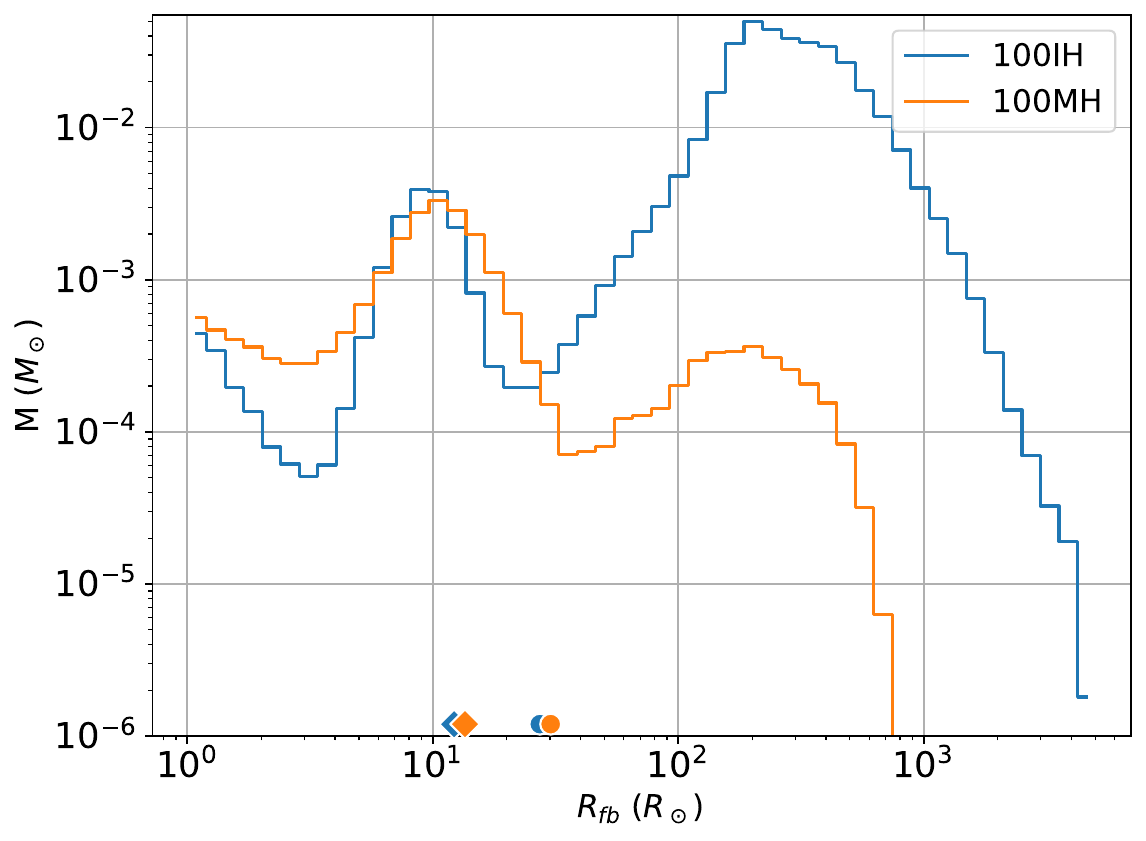}
    \caption{Histograms of bound mass as a function of fall-back radii for $q= 0.68$ (top), $q= 0.85$ (middle), and $q= 1$ (bottom), where the blue and orange lines are the high and tabulated EoS simulations for each value of $q$, respectively. Plots are generated at $t=t_{\rm end}$. The circles and diamonds in each plot represent the orbital distance from the centre of mass of the primary and companion cores, respectively. 
    }
    \label{fig:fallback_radii}
\end{figure}

Figure~\ref{fig:fallback_radii} shows the distribution of the bound gas with respect to its fall-back radius, as given by Equation~\ref{eq:Rc}. The blue lines are for the IH simulations, that unbind only part of the gas, leading to more massive discs, while the orange lines are for the MH simulations that unbind most of the gas leading to a significantly less massive disc. We find that the distribution of fall-back material is narrower for the tabulated EoS simulations and depends somewhat on the mass ratio of the binary, with increasing $q$ leading to slightly larger inner disc's edge. This is due to simulations with larger values of $q$  having more angular momentum because they start mass transfer when the companion is farther out. Our simulation with the lowest mass ratio (68IH/MH) results in a fall-back disc with a radial span of 10--3000~\rs\ and 20--750~\rs\ for the ideal gas and tabulated EoS simulations, respectively. The secondary, smaller peak near 3--10~\rs, is due to the gas that effectively remains bound around the stellar cores (seen in  Figure~\ref{fig:68_fb_timescale} as the two upside-down triangle shapes at small values of $r$). 
\rev{For the 85 and 100 simulations similar values are observed (15--4500~\rs\ and 35--750~\rs\ for the 85IH and 85MH simulations, respectively and 20--4500~\rs\ and 40--750~\rs\ for the 100IH and 100MH simulations respectively.)}

Using this range of fall-back radii we now calculate an approximate viscous timescale from Equation~\ref{eq:tVisc}. The temperature of the gas surrounding our post-interaction binary is in the range $T \approx 4\times10^3 - 1\times10^4$~K for all simulations. This then corresponds to a sound speed in the range of $c_{\rm s} = 5 -12$~km~s$^{-1}$. If we take $\alpha = 0.01-0.1$, then the viscous timescale, and thus the expected circularisation timescale of our circumbinary disc is $\tau_{\rm visc}\approx10-100$ years.

Table~\ref{tab:fb_summary} summarises the amount of mass expected to fall back onto the binary (the bound mass at the end of the simulations) for our  simulations. In the case of the tabulated EoS simulations, fall-back masses are in line with the observed masses of post-giant circumbinary discs (0.02-0.05~\ms; \citealt{Pourmand2025}). How much more massive these discs may have been at the time of formation is unclear, but with the observed mass range in mind, we can state that only up to  $\sim5$\% of our envelope would need to remain bound. In Table~\ref{tab:fb_summary} we also list the average location, $R_{\rm fb} \sim R_{\rm c}$, of this mass from the centre of mass of the system (binary plus bound gas), and the distances from the centre of mass of each of the two stars in the binary, $d_{\rm orb,1/2}$.  

\begin{table}
    \centering
    \begin{tabular}{lclll}
    \hline
         Sim.&  <$t_{\rm fb}$>   &$M_{\rm fb, end}$  &$R_{\rm fb,end}$ &$d_{\rm orb,1/2}$\\
 Name& (yrs)&(\ms) &(\rs)&(\rs)\\
         \hline

 68IH& 170&0.30 & 100, 240&11, 7\\
 68MH& 360&0.020 & 140&13, 9\\
 85IH& 720&0.32 & 340&17, 9\\
 85MH& 420&0.0086 & 200&21, 11\\
 100H& 1100&0.34 & 200&27, 12\\
 100MH& 80&0.0031 & 200&30, 14\\

 \hline
 \end{tabular}
    \caption{Mass-weighted average fall-back times (Equation~\ref{eq:t_fb}) and total mass of the bound envelope at end of inspiral ($t=t_{\rm end}$) for each simulation. The total mass of the fall-back material is calculated from the area underneath the curve for the regions exterior to the orbit, i.e. the outer set of peaks for each simulation in Figure~\ref{fig:fallback_radii}.  $R_{\rm fb}$ is the disc radius from the centre of mass, (the peak of each histogram in Figure~\ref{fig:fallback_radii}, note for the peak plateau in the 68IH simulation we have marked the inner and outer radii), and the distance from the centre of mass of the core and companion  respectively ($d_{\rm orb,1/2}$),  is given to show the location of the disc with respect to the orbit of the binary (the circles and diamonds in Figure~\ref{fig:fallback_radii}, respectively).}
    \label{tab:fb_summary}
\end{table}

\section{Discussion}
\label{sec:Discussion}

In this section we discuss the impact of the mass ratio on common envelope interactions between RGB stars and compact companions. \rev{The CE interaction likely plays an important role in the evolution of binaries with periods less than approximately 2000 days \citep{Krynski2025}. As such we discuss the weakening of the CE interaction} in the context of post-RGB (and post-AGB) binaries. In particular we consider the final orbital separation, and whether a stable circimbinary disc can form.

\subsection{Mass Transfer Stability and Final Orbital Separation}

Our simulations demonstrate that the duration of the stable mass transfer phase preceding the CE inspiral is sensitive to the mass ratio $q = M_2/M_1$, the resolution, and the equation of state (EoS). For example, ideal gas EoS simulations 68IH, 85IH and 100IH, with increasing $q$ see the pre-inspiral time go from 39 to 64 to 72 years (shorter, but with the same trend for the tabulated EoS simulations; see $t_i$ in Table~\ref{tab:CE_summary}). Noting that the pre-inspiral time is not converged, such that this value must be a lower limit, our simulations support the analytic prediction \citep{Tout1991,Frank2002,Dan2011} that mass transfer begins to stabilise for $q\gtrsim1$. 

Mass transfer rate through $L_1$ early in the simulation has a slight correlation with $q$ ($\dot{M}_{L_1,i} = 4\ {\rm to}\ 9 \times 10^{-4}$~\ms~yr$^{-1}$ for the ideal gas and tabulated EoS simulations, similarly), but this is not a converged quantity, with the low resolution simulations showing values that are about twice those of the respective high resolution simulations. The mass transfer rate remains approximately constant during the pre-inspiral time for the tabulated EoS simulations. A longer the pre-inspiral mass loss phase achieved by increasing resolution may not increase the total amount of mass transferred as the mass transfer rate also decreases with increased resolution. It is therefore possible that the lack of convergence of the length of time before inspiral may not change the amount of total mass transferred before inspiral.

The final separation depends on the mass ratio with values of 27, 38 and 49~\rs\ for the high resolution, ideal gas EoS 68IH, 85IH and 100IH simulations at the end of the inspiral, with slightly decreasing values afterwards. The final separation after the inspiral is converged and it is also very similar for the different EoS. 
Contributing factors to this trend are both the higher initial angular momentum of the higher $q$ simulations (due to the larger initial orbital separations and mass) {\it and} the larger orbital energy inherent to the more massive companions (as observed by \citet{Passy2012}, who started different $q$ simulations at the same initial orbital separation).  This said, the largest value, at $\sim 50$~\rs\ (or a period of $\sim 40 $ days) is still smaller than the smallest separation of the post-RGB/AGB binaries ($\sim 100$~\rs; or a period of $\sim 100$ days; \citealt{kluska_population_2022}), {\it showing that entering the CE will tend to cause too much orbital shrinkage to explain the observed separations, no matter what the mass ratio}. 

The $q = 1.5$ is not representative of post-RGB binaries because the companion could not be so much more massive than the primary and still be on the main sequence. (These massive companions could, however, represent massive WDs). Our 150IL/IH/MH simulations allowed us to investigate an even more extreme value of $q$. Our ideal gas simulation, 150IH, shows limited inspiral within the limited simulation runtime, and the orbital separation begins to plateau along with the mass transfer rate itself. The tabulated EoS simulation, 150MH has a modest, shallow inspiral, leading to a final separation at the end of the simulation of $\sim 140$~\rs\ (a period of $\sim 130$~days) at the low end of the observed separations (plateauing but still decreasing gently). This confirms the trend and suggests that, in real systems where mass transfer proceeds more gradually than in simulations, higher $q$ values likely allow the binary to avoid inspiral and achieve wider post-interaction separations \rev{that are more inline with observed post-giant systems}.

With increasingly stable mass transfer enabled by larger values of $q$, heat loss from the surface of the star may be sufficient to stabilise and slow down the mass transfer rate and possibly avoid the inspiral altogether. The stellar envelope could be removed in this way. Alternatively, the binary could widen and mass transfer could stop altogether. In such case, the RGB star envelope removal would eventually complete by regular winds, leaving a post-RGB binary systems with uncertain orbital parameters. 

\subsection{Unbinding the Envelope}
\label{ssec:unbinding-the-envelope}

Higher resolution simulations typically unbind less mass than their lower-resolution counterparts unless aided by a tabulated EoS that includes recombination energy. This agrees with findings by \citet{Reichardt2020} and \citet{gonzalez-bolivar_common_2022} and emphasizes the need for caution when interpreting the total unbound mass in simulations. While the use of a tabulated EoS has become common practice in CE simulations, the retention of all energy by our adiabatic simulations means that the unbinding observed with tabulated EoS must be considered an upper limit. Hence, when discussing the circumbinary disc mass, the ideal gas simulation, which unbind less mass, would generate a more massive disc, while the tabulated EoS simulations, which unbind almost the entire envelope, would generate a much lower mass disc. 

Our 150MH simulation illustrates that nearly the entire envelope can be unbound without a deep dynamical inspiral. This suggests that, in physical systems, a slow process of unbinding enabled by recombination energy—combined with more stable mass transfer—could be sufficient to eject the envelope.

\subsection{Post-Interaction Circumbinary disc Formation}
\label{ssec:post-interaction_disc_formation}

There are at least two ways to form a circumbinary disc: via $L_2$ and $L_3$ mass outflow and by fallback of bound mass after a CE ejection. In the current context of CE interactions, the former method works if the CE ejection that follows the formation of a circumbinary disc via $L_2/L_3$ does not sweep the disc away. The latter method works if there is sufficient bound mass, with sufficient angular momentum to form a disc around the binary.

We observe almost no increase in the mass outside $L_2$ by the start of the inspiral as a function of $q$ for all simulations. Of the mass outside $L_2$, only about half is still bound at that time for all simulations. That is the mass that would be destined to form a disc. The 150IH/MH simulations shows that pushing $q$ to higher values increases the amount of mass lost through $L_2$ only for the MH simulation, but for all 150 simulations it increases the fraction of that mass that is unbound, leaving almost no gas to make a disc. This, once again, argues {\it against} the hypothesis by \citet{Reichardt2019} that higher $q$ simulations could form a more massive (bound) disc. 

This said, some gas (whether ejected via $L_2$ before the inspiral, or ejected later on via the CE ejection) remains bound to the system (more so for the ideal gas EoS simulations and less for the tabulated EoS ones, as explained in Section~\ref{ssec:unbinding-the-envelope}).  By modeling bound gas as ballistic particles, we find fall-back timescales of centuries to about one millennium. The fall-back/circularization radii range from 0.5 to about 1.5~au, large enough to be stable around the binary, but much smaller than the discs observed around post-RGB/AGB binaries. It is possible that viscous spreading will take place on relatively short timescales thereafter, resulting in larger discs.

\subsection{Circumbinary material and orbital elements}

The fall back of material that remains bound to the system post-inspiral presents a potential opportunity for \textit{secondary interactions} that may alter the orbital elements. In Section~\ref{ssec:post-interaction_disc_formation} we discussed how the fall back of material {\it in these simulations} should form a disc outside the orbit of our binaries. This said, some ejected material may still return close enough to the orbit of the binary such that it may re-accrete before the disc formation\footnote{\citet{Kuruwita2016} showed that if insufficient mass is unbound in the CE, the substantial fallback would lead to renewed orbital interaction leading to additional orbital shrinkage.}. 

Even if some gas had so little angular momentum that it would fall back onto the orbit, it is unlikely that an interaction between relatively little gas and the binary would lead to a significant decrease of the binary's orbital separation: at orbital separations of $\sim 50$~\rs, the energy required to shrink the orbit by half, down to 25~\rs, is approximately the same amount of energy it takes to shrink the orbit  from the original $\sim$200~\rs\ to $\sim50$~\rs. This said, any gas that falls back to within the binary orbit should continue to interact with the binary through both accretion and resonances, until it is redistributed leaving a cavity of 2--3 times the orbital separation  \citep{artymowicz_dynamics_1994}. We will consider this phase of disc adjustment in a future work.

\citet{Gagnier2023,Gagnier2024,Gagnier2025} have investigated the post interaction environment using 3D hydrodynamic simulations. Using a setup similar to \citet{Morris2006,Morris2007,Morris2009} and \citet{Hirai2021} they initialised their systems with a distribution of circumbinary material endowed with angular momentum consistent with that expected from material spun up during the interaction. They found that the material forms a thick, bound, out-flowing disc in the equatorial plane, with inward falling material around the poles. This orbiting material then grows eccentric with local over-densities present despite the central binary's orbit being held circular \citep{Gagnier2023}. It is speculated this eccentric material could influence the eccentricity of the binary, which is interesting since many observed post-RGB/AGB binaries feature non-zero orbital eccentricities. We suspect, however, that tides, particularly after the CE, are much too efficient at circularising the orbit, such that binary-gas interactions are unlikely to meaningfully change this. This does still leave room for the decay of the orbital separation even after the inspiral, which \citet{Gagnier2023} found may be possible on timescales of $10^3-10^4 P_{\rm orb}$ in the presence of accretion. The presence of magnetic fields also seems to have little effect on the future binary evolution, though can act to modestly shape the resulting disc-like structure \citep{Gagnier2024}. 

Previous simulations \citep{Reichardt2019,gonzalez-bolivar_common_2022} similar to ours have also shown that material surrounding the binary is typically dragged into co-rotation which is in part responsible for halting the inspiral due to a quenching of net torques. Recently \citet{Gagnier2025} similarly found that any further evolution of the binary in the post-interaction environment is likely due to interactions between the inner, co-rotating layers around the binary and those outside this region that are not co-rotating. However, as previously mentioned, the resulting influence on the binary's orbit is expected to be minimal.

\section{Conclusions and Summary}\label{Sec:Conc}

We performed a suite of 3D hydrodynamic simulations of CE evolution between a 0.88~\ms\ RGB star and compact companions with varying mass ratios, $q \equiv M_2/M_1 = 0.68$--$1.5$ and two equations of state, to investigate regimes of increasing mass-transfer stability and their impact on the CE interaction outcome. We find that: 

\begin{itemize}
\item Increasing $q$ extends the duration of the pre-inspiral mass transfer phase, weakens the inspiral, and leads to wider final orbital separations. Some of these quantities, with the exception of final separation, are not converged with respect to simulation resolution. However, increasing the resolution meaningfully is excessively expensive computationally. 
\item We expect that binary interactions with high $q$ values would be even more stable in nature, with even longer pre-inspiral timescales, although the final separation would be similar to the one obtained here. 
\item For sufficiently high $q$, the classical CE inspiral might be entirely avoided, suggesting that the envelope of the donor could be unbound through sustained, stable mass loss rather than dynamical orbital decay. 
\item Final orbital separations, even for large $q$ values are of the order of 50--100~\rs, or periods of 30--120 days, still generally smaller than the range of semi-major axes and periods of post-RGB and post-AGB binaries.
\item We do not witness the formation of a post-CE, circumbinary disc  through mass ejection via $L_2$. Tabulated equation of state simulations unbind all the $L_2$ ejecta, while ideal gas simulations result in only about half of the $L_2$ ejecta remaining bound to the system. In both cases, however, the mass ejected during and after the inspiral, tends to sweep away any low mass disc that may have formed just before. 
\item \rev{In all simulations a fraction of the envelope remains bound to the system and we calculate that it will fall back towards the binary forming a disc that spans 15--4000~\rs\ (0.1 -- 20~au) for ideal gas simulations, and 25--750~\rs\ (0.1 -- 3.5~au), for tabulated EoS simulations, in a time frame between a tenth of a year and 1000 years, with the bulk of the bound mass returning within a year. A further 10-100 years may be needed to viscously spread the disc further. } Such discs may develop characteristics that are in line to those surrounding  post-RGB/AGB binaries. 
\item Our findings suggest that high mass ratios at the time of the interaction may help to provide ideal conditions for a weakened CE that results in wider binaries with  circumbinary discs; however, this alone is unlikely to be the solution to the formation of such observed systems. In a second paper, we will investigate the effect of mass ratio on AGB, instead of RGB stars which, with their lighter envelope may exhibit even weaker interactions. 
\item Finally, we start to see a behaviour that is critically different from a CE inspiral, only for our $q = 1.5$ simulation. This ratio may be reduced if we could simulate a more realistic stellar surface, where rapid cooling may stabilise the mass transfer and avoid the CE for lower values of $q$. 
\end{itemize}

\paragraph{Acknowledgments}
This work was supported by resources awarded under Astronomy Australia Ltd's ASTAC merit allocation scheme on the OzSTAR national facility at Swinburne University of Technology. The OzSTAR program receives funding in part from the Astronomy National Collaborative Research Infrastructure Strategy (NCRIS) allocation provided by the Australian Government, and from the Victorian Higher Education State Investment Fund (VHESIF) provided by the Victorian Government. This research was supported by the Commonwealth through an Australian Government Research Training Program Scholarship [DOI: https://doi.org/10.82133/C42F-K220]. LS is research director at the F.R.S-FNRS. \rev{We acknowledge discussions with Mike Lau, Taissa Danilovich, Kayla Martin, Ana Juarez, Stephen Neilson, Chunliang Mu, and Ali Pourmand. }






\bibliography{references,references_orsola,references_orsola_2}

\appendix 
\renewcommand{\thefigure}{A\arabic{figure}} 
\setcounter{figure}{0}
\section{Resolution-dependent unbinding for low resolution simulations}
\label{sec:appen1}

\begin{figure*} 
\centering
    \includegraphics[width=0.48\linewidth]{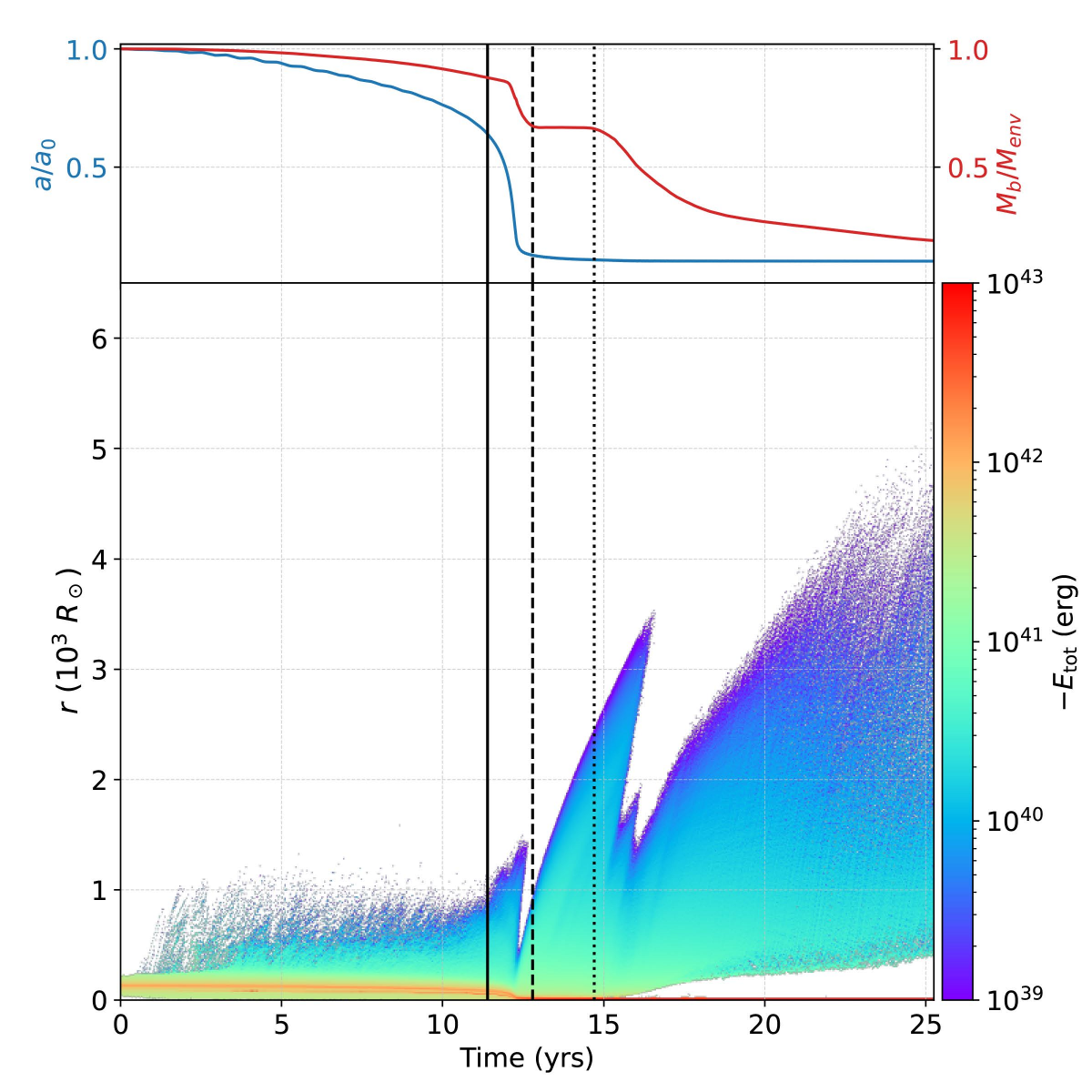}
    \includegraphics[width=0.48\linewidth]{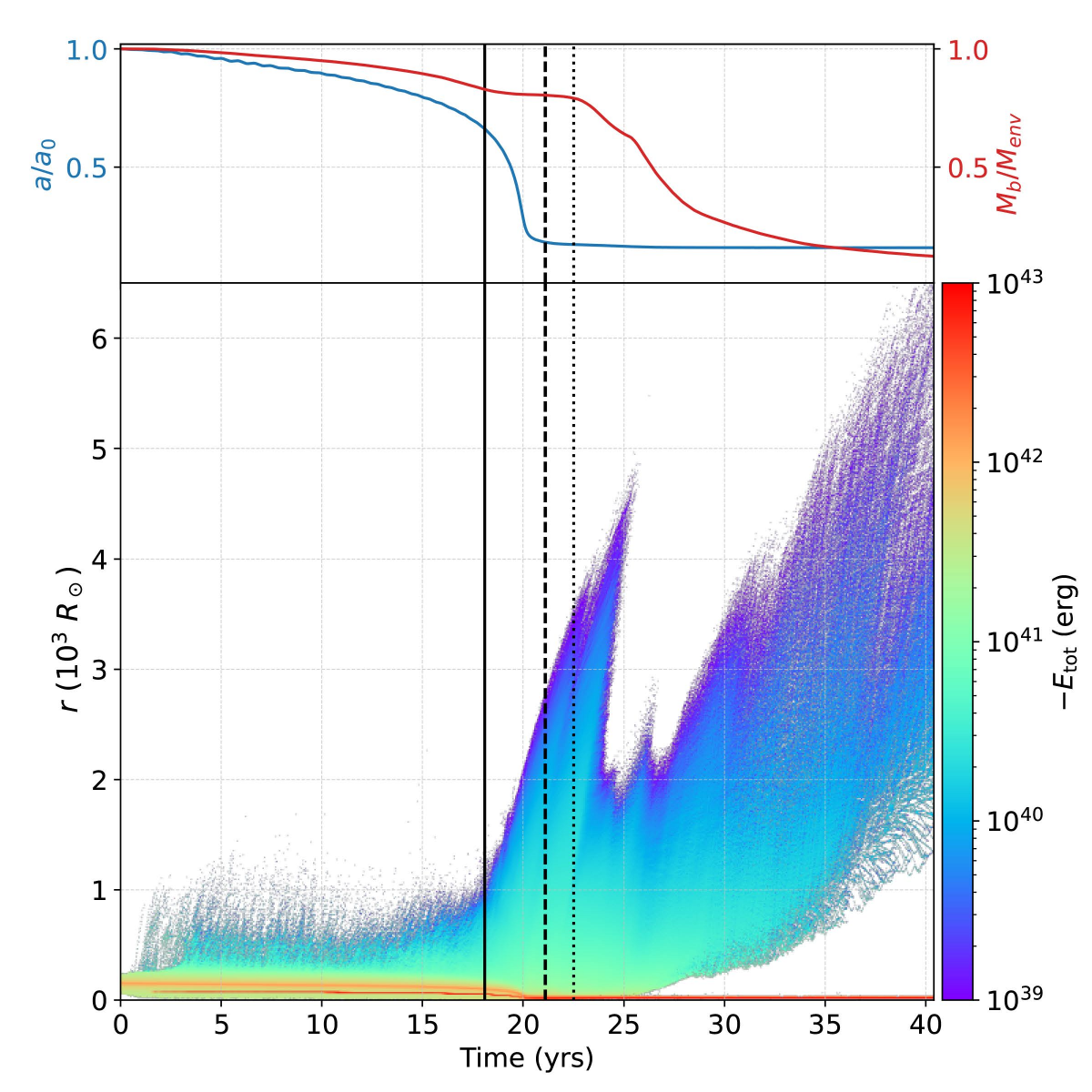} 
    \includegraphics[width=0.48\linewidth]{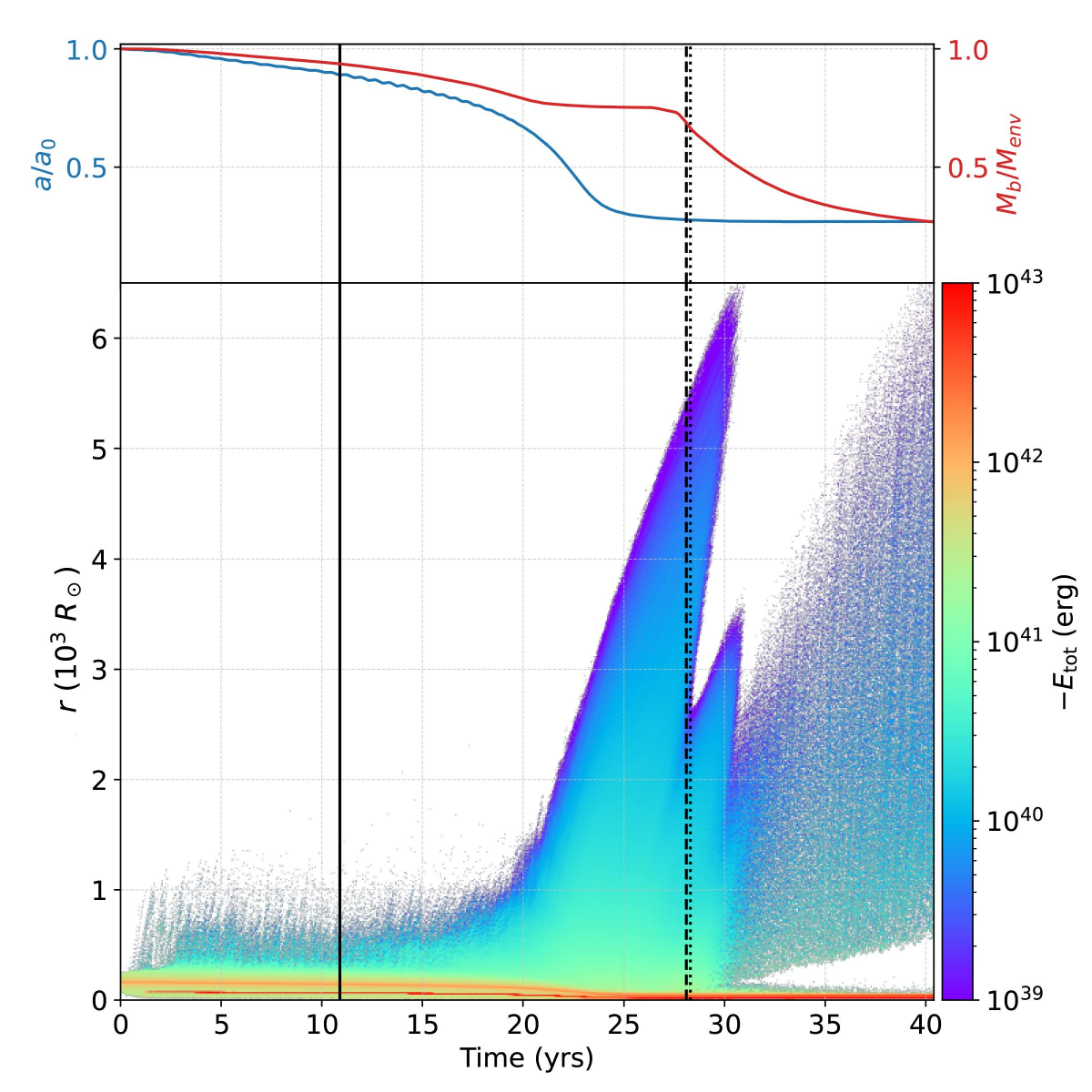}
    \includegraphics[width=0.48\linewidth]{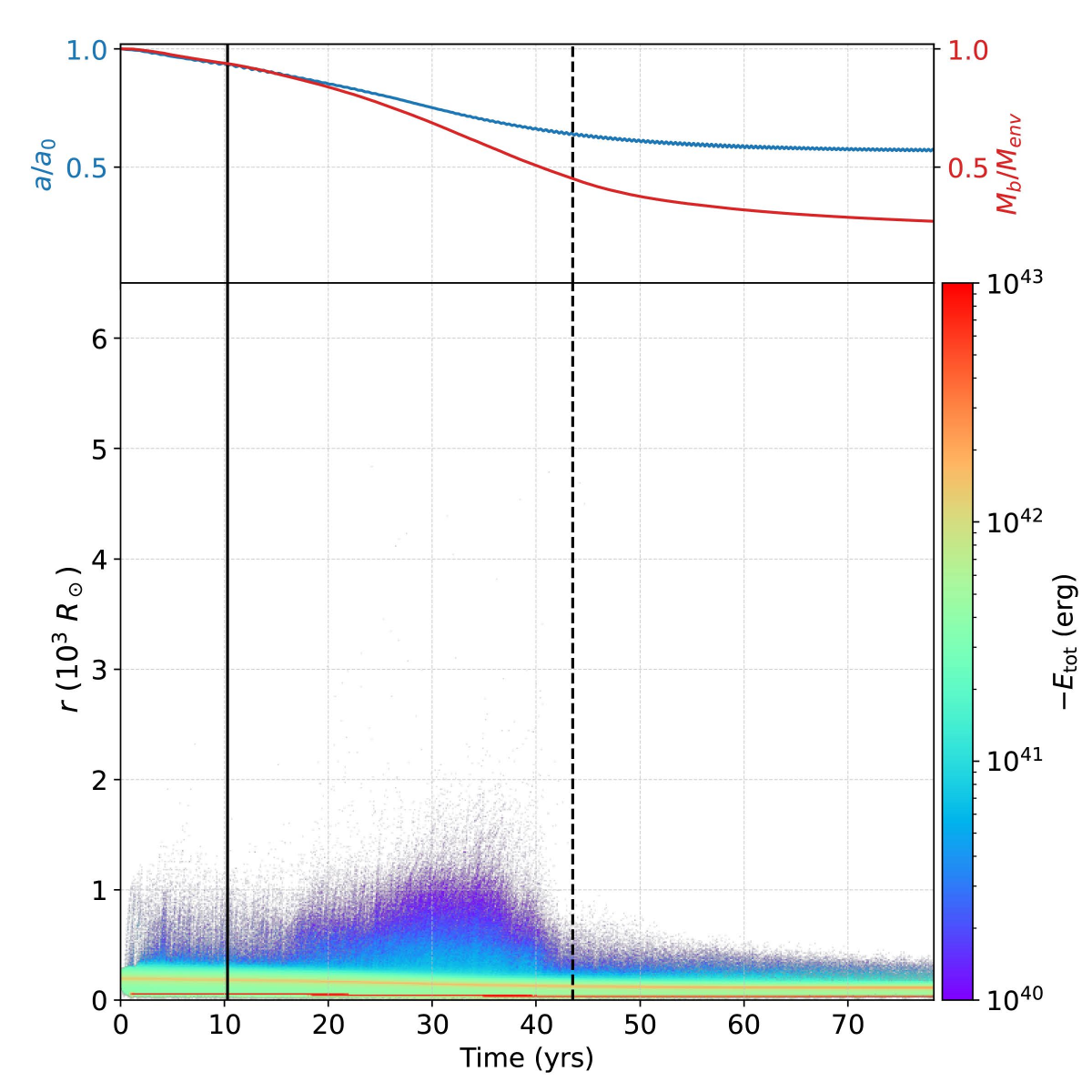} 
    \caption{Distribution of bound gas for low resolution  simulations 68L (top left), 85IL (top right), 100IL (bottom left), and 150IL (bottom right). A demonstration of resolution-dependent gas unbinding at the base of the envelope is present in all simulations that undergo a strong inspiral.}
    \label{fig:allLH_bound_inspiral}
\end{figure*}

In Figure~\ref{fig:allLH_bound_inspiral} we show the distribution of bound gas in ideal gas EoS simulations demonstrating that the low resolution simulations unbind gas at the base of the envelope (the white zone that develops after the dotted vertical lines in three of the panels in Figure~\ref{fig:allLH_bound_inspiral}), a behaviour typical of the resolution-dependent unbinding discussed by \citet{gonzalez-bolivar_common_2022}. This is not observed in high resolution simulations (see Figure~\ref{fig:LH_bound_inspiral}). Interestingly, simulation 150IL does not display the wave of unbinding at the base of the envelope. By the time we stop that simulation, there has been only  little, gentle inspiral, which however has unbound 73\% of the envelope mass (Table~\ref{tab:L2_b_ub_data}). Yet the modality of this unbound mass is clearly distinct from that of the other three simulations in Figure~\ref{fig:allLH_bound_inspiral}.

\section{On the comparison of numerical and analytical mass transfer rates}
\label{sec:appendixB}
\begin{figure*}
    \centering
    \includegraphics[width=\linewidth]{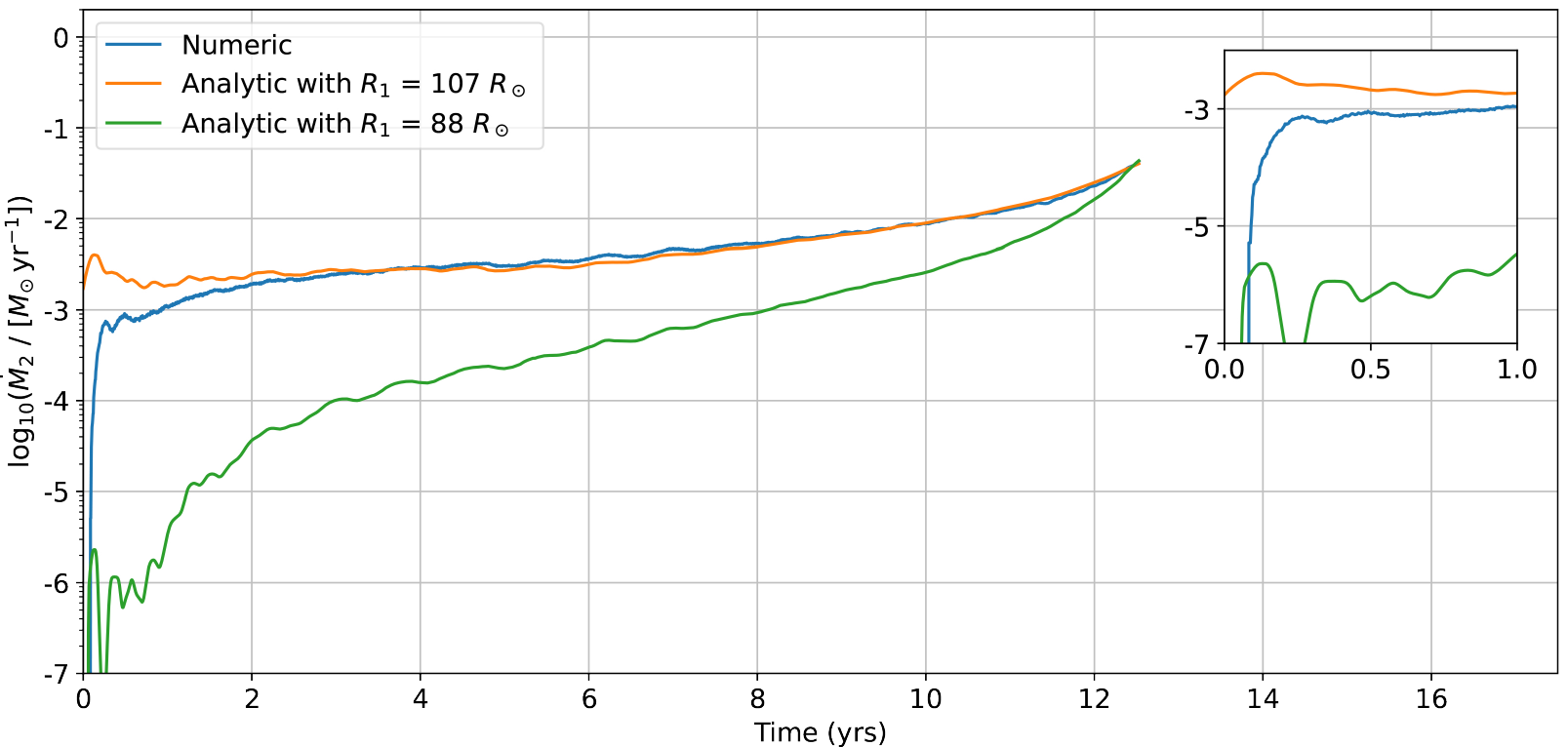}
    \caption{A recreation of figure 7 of \citet{Reichardt2019}. The orange and blue lines are those used in that work, representing the mass transfer rate calculated using the analytical equation for mass transfer, along with quantities measured using the simulation (orange line) and the mass transfer measured directly from the simulation (blue line). The green line shows the mass transfer rate calculated analytically, but this time using a somewhat smaller value for the stellar radius, which is however still a reasonable estimate, resulting in a significantly lower rate of mass transfer and a mismatch between analytical theory and the simulation.}
    \label{fig:reich_mdot}
\end{figure*}

\citet{Reichardt2019} carried out a comparison between the mass transfer rate before the CE inspiral in their simulations, versus values derived using the analytical approximation of \citet{PacSien1972}, using variables measured from the simulations. In that work the comparison was satisfactory, but, we argue here, it was also fortuitous and due to the specific value of the stellar radius they measured from simulations, which should have been considered very uncertain.

In Figure~\ref{fig:reich_mdot} we show a recreation of their figure 7, where their numerical mass transfer rate was compared with the analytical prescription using a stellar radius of 107~\rs. This value was measured from the core of the primary in a direction perpendicular to the orbital plane (the direction with the least amount of distortion due to the binary interaction). The radius of the star in that direction was defined as the distance between the core and the least dense particles, then adding twice their SPH smoothing lengths as has been standard practice for SPH simulations \citep[e.g.,][]{Nandez2014}. The radius value is used in the analytical expression to determine by how much the stellar radius exceeds the $L_1$ point. This `excess' is then taken to the {\it third power} to determine, along with other variables, the mass transfer rate. As it happens the radius value determined leads to an analytical mass transfer rate that closely matches the numerical mass transfer rate! However, a radius value as little as 20\% smaller (88~\rs) changes the fit considerably (Figure~\ref{fig:reich_mdot}). This goes to show that when a satisfactory answer is found, one tends not to investigate matters further.

Given the difficulty of defining radii in SPH even for spherical stars, and given the complexity of measuring a distorted star, we conclude that it is uninstructive to carry out this comparison.

\section{Energy and angular momentum conservation}
\label{sec:appendixC}
\begin{figure*}
    \centering
    \includegraphics[width=\linewidth]{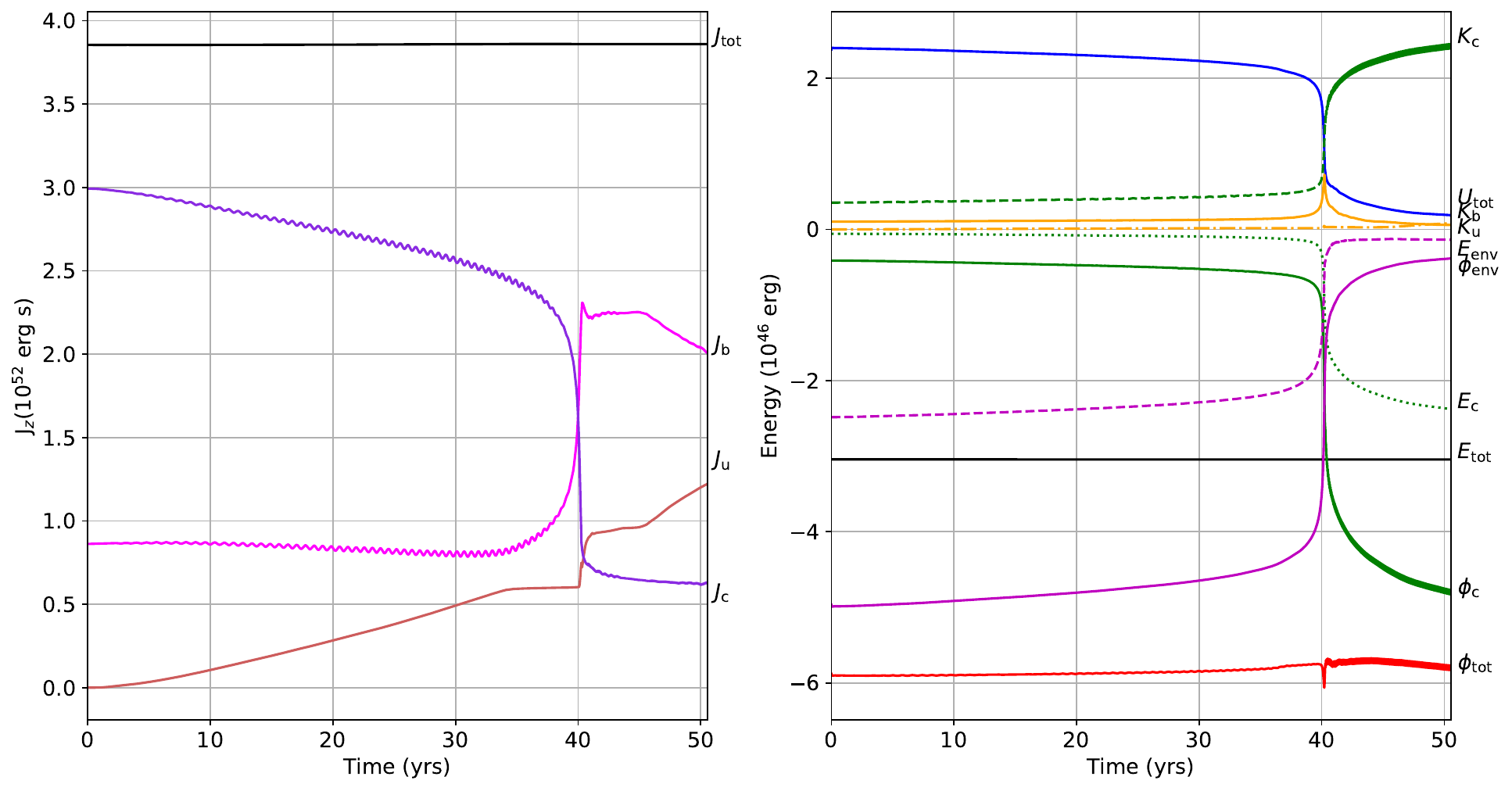}
    \caption{Plots of angular momentum (left) and energy (right) evolution for the 68H simulation, as an example. In each plot the subscripts $tot$, $b$, $u$, $c$, $env$, refer to the total, bound, unbound, cores, and gas respectively. In the plot of energy, $K$, $U$, and $\phi$, are the kinetic energy, thermal energy, and the potential energy respectively, while $E$ is used for the combination of potential, kinetic and thermal as appropriate to the cores, envelope or total.} \label{fig:conservation}
\end{figure*}

Energy and angular momentum are conserved to excellent precision in all of our simulations. Figure~\ref{fig:conservation} shows the 68IH simulation as an example. In all of our simulations momentum is conserved to within 0.2\%, and total energy is conserved to within 0.1\%.





\end{document}